\documentclass[superscriptaddress,aps,journal=prl,preprint]{revtex4-2}
\usepackage{CJK}
\usepackage{graphicx}
\usepackage{bm,amsmath,amssymb}
\usepackage{color}
\usepackage{dcolumn}
\usepackage[colorlinks=true, linkcolor=blue, citecolor=blue, urlcolor=blue]{hyperref}

\begin{document}
\begin{CJK*}{UTF8}{gbsn}

\title{Double Helix of Atomic Displacements in Ferroelectric PbTiO$_3$}
\author{Yihao Hu (胡逸豪)}
\email{huyihao@westlake.edu.cn}
\affiliation{Zhejiang University, Hangzhou, Zhejiang 310058, China}
\affiliation{Department of Physics, School of Science, Westlake University, Hangzhou, Zhejiang 310030, China}

\author{Shi Liu (刘仕)}
\email{liushi@westlake.edu.cn}
\affiliation{Department of Physics, School of Science, Westlake University, Hangzhou, Zhejiang 310030, China}
\affiliation{Institute of Natural Sciences, Westlake Institute for Advanced Study, Hangzhou, Zhejiang 310024, China}

\begin{abstract}

Recent theoretical work has predicted the existence of a ``dipole spiral" structure in strained freestanding membranes of PbTiO$_3$, suggesting a potential route to enhanced electromechanical responses [\href{https://journals.aps.org/prl/abstract/10.1103/PhysRevLett.133.046802}{PRL \textbf{133}, 046802 (2024)}]. However, its microscopic nature, energetic landscape, and electronic properties remain largely unexplored from a first-principles perspective.
Here, using density function theory on PbTiO$_3$ under biaxial tensile strain, we identify a novel form of polar order: a chiral, non-collinear ferroelectric double helix. We find that the Pb- and Ti-cation sublattices form two distinct, intertwined helices, reminiscent of DNA. This topology is stabilized by a collective helical twisting of the oxygen octahedral framework, which gives rise to an electric Dzyaloshinskii-Moriya-like interaction. The resulting structure, which can be canceptualized as a ``self-Moir\'e" crystal, exhibits two coupled functionalities. First, it possesses a rotational pseudo-zero-energy mode that underpins a giant piezoelectric response ($e_{33}\approx$16 C/m$^2$). Second, the long-period potential reconstructs the electronic band structure, leading to a multi-valley electronic topology at the valence band edge. Our work establishes a physical route to designing complex chiral order that supports both giant electromechanical coupling and multi-valley electronics.

\end{abstract}

\maketitle
\end{CJK*}

\newpage

The past decade has witnessed a surge of interest in topological polar structures within ferroelectric oxides. Real-space textures such as polar vortices~\cite{Rodriguez09p1127,Balke12p81,Yadav16p198}, skyrmions~\cite{Das19p368,Han22p63}, and merons~\cite{Wang20p881}, analogous to their magnetic counterparts, have been realized.
These discoveries have demonstrated that by carefully balancing elastic, electrostatic, and polarization-discontinuity gradient energies through epitaxial strain and interfacial engineering~\cite{Ramesh19p257}, one can overcome the strong crystalline anisotropy of ferroelectrics and stabilize complex, non-collinear polarization patterns. Helical structures are ubiquitous in nature, manifesting across all length scales. From the spiral arms of galaxies to the turbulent vortices of weather systems, down to the intricate spin helices that drive multiferroicity in quantum materials~\cite{Gross17p252,Sando13p641}, this chiral form represents an elegant solution to complex energetic constraints. Perhaps most iconically, the double helix of DNA encodes the fundamental blueprint of life itself. A compelling question thus arises in condensed matter physics: can a similar helical order be realized for electric dipoles within a crystalline solid, and what new physics would such a ``ferroelectric helix'' unlock?

The realization of helical order arising from competing interactions is a well-established concept in modern magnetism, but achieving an analogous helical texture of electric dipoles has remained challenging. One notable development was the experimental observation of an incommensurate helical dipole texture in the chemically doped quadruple perovskite BiCu$_{0.1}$Mn$_{6.9}$O$_{12}$, which was shown to be stabilized by an intrinsic competition between lone-pair activity and orbital ordering~\cite{Khalyavin20p680}.
In parallel, theoretical work has proposed a distinct route toward ferroelectric helicity: stabilizing a ``dipole spiral'' in the archetypal ferroelectric PbTiO$_3$ under large in-plane biaxial tensile strain~\cite{Hu24p046802}. Unlike the very small out-of-plane polarization ($<$ 20~$\mu$C/m$^2$) in BiCu$_{0.1}$Mn$_{6.9}$O$_{12}$, this phase in PbTiO$_3$ exhibits a stable out-of-plane ferroelectric polarization comparable to that of the conventional tetragonal ($T$) phase (0.7~C/m$^2$), making it a potential example of a ferroelectric helix.
Moreover, this topological structure has been associated with a potentially enhanced piezoelectric response, pointing to a mechanism for electromechanical coupling that may be rooted in the underlying topology. These findings motivate further investigation into the microscopic origin and functional implications of such a phase.
Despite recent interest, a comprehensive understanding of this helical phase from a first-principles perspective remains incomplete. Key questions persist regarding its microscopic structure, energetic stability relative to conventional ferroelectric phases, the origin of its functional properties, and the extent to which it modifies the material's electronic behavior.

Here, we employ density functional theory (DFT) calculations to investigate the fundamental physics of the emergent phase in strained PbTiO$_3$. Our results indicate that the dipole spiral corresponds to a ferroelectric double helix. 
Specifically, we find that the Pb and Ti cation sublattices form two distinct, intertwined helical paths with a stable, non-parallel phase relationship between them. The stability of this configuration can be traced to a collective helical twisting of the oxygen octahedral framework, which gives rise to an electric Dzyaloshinskii-Moriya-like interaction (eDMI)~\cite{Dzyaloshinsky1958p241,Zhao21p341}.
This chiral structure, which may be viewed as a one-dimensional ``self-Moir\'e" crystal, exhibits a rotational pseudo-zero-energy mode that contributes to its enhanced piezoelectric response. Furthermore, we find that the structural topology leads to a reconstruction of the electronic band structure, resulting in a multi-valley electronic character at the valence band edge.
Overall, our work outlines a theoretical framework that links structural chirality, emergent eDMI, electromechanical response, and electronic band topology, offering insights into potential design strategies for multifunctional polar materials.

\begin{figure*}
    \centering
    \includegraphics[width=1\textwidth]{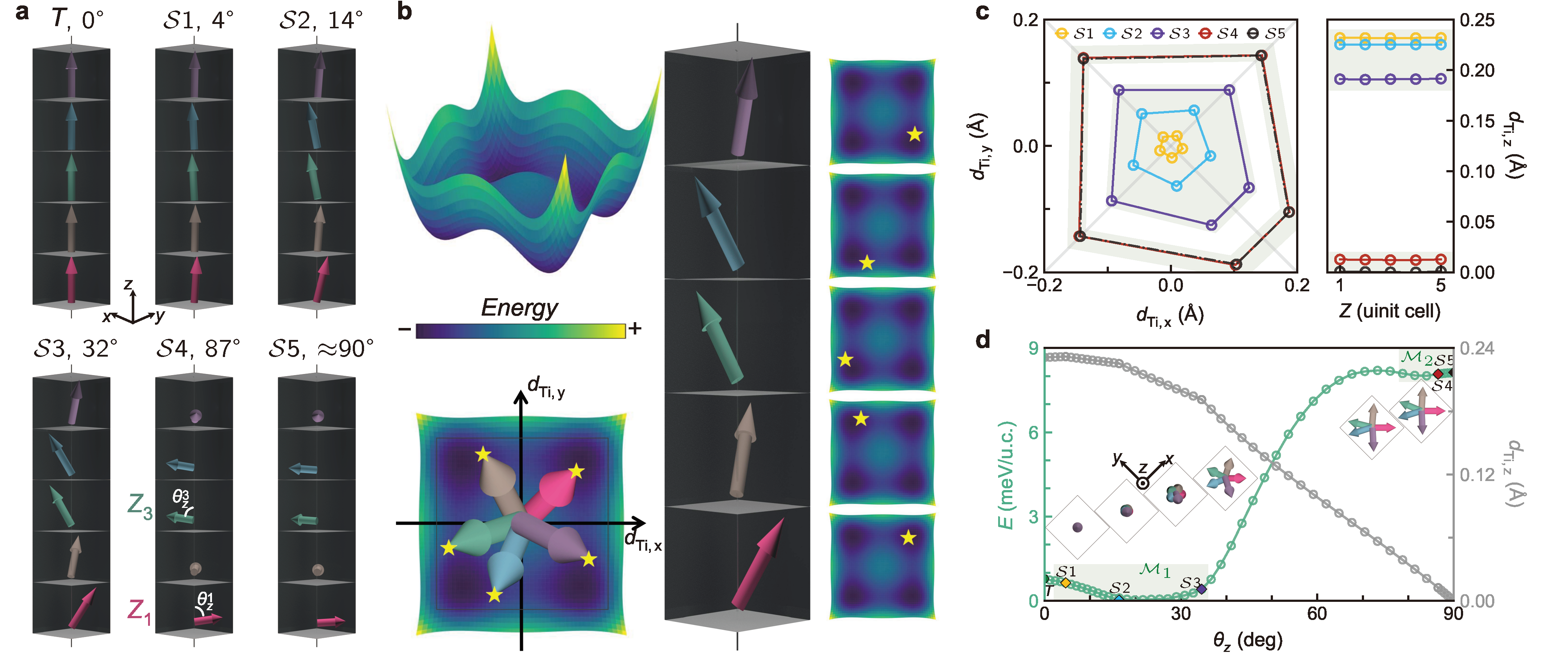}
    \caption{\textbf{A Family of Dipole helical phases and Their Energetic Landscape.} 
    (a) Schematics of $T$ and stable dipole helical phases ($\mathcal{S}$1-$\mathcal{S}$5) in strained ($a=b=3.970$~\AA) PbTiO$_3$, distinguished by their polarization tilt angle $\theta_z=\frac{1}{N}\sum_{k=1}^N \theta_z^k$. 
    (b) The quadruple-well potential energy surface arising from the $C_4$ symmetry of the strained lattice. The yellow stars represent the in-plane polarization vectors ($d_{\text{Ti},xy}$) for each of the five layers of a representative helical phase, illustrating how the polarization is energetically favored to lie within the potential wells.
    (c) The left panel shows the top-down view of dipole spiral's in-plane polarization path, formed by connecting the vertices (the yellow stars) from (b). This path is a deformed polygon, a direct consequence of the pinning effect of the quadruple-well potential. The right panel plots the out-of-plane polarization component ($d_{\text{Ti},z}$) as a function of the layer index ($Z$).
    (d) Energy and $d_{\text{Ti},z}$ landscape along a linear interpolation path between $T$-$\mathcal{S}$1-$\mathcal{S}$5. The insets are the top view of the schematics in (a), corresponding to the left panel of (b).
    }
    \label{Fig1}
\end{figure*}

Our first-principles calculations for PbTiO$_3$ under substantial biaxial tensile strain indicate the stabilization of a ground state that differs from conventional ferroelectric phases (see Supplemental Material (SM)~\cite{SIyhh}, Sec.\MakeUppercase{\romannumeral 4}). The system adopts a set of chiral dipole spirals, topological polar structures that bear analogy to spin spirals in magnetic materials~\cite{Tsunoda89p10427,Tsunoda93p133}. 
A key geometric feature of this phase is its helical character, which we quantify using the Ti-cation displacement $d_{\rm Ti}$ as a proxy for local polarization~\cite{KingSmith93p1651}. Specifically, the dipoles under 2.3\% strain ($a=b=3.970$~\AA) in a 1$\times$1$\times$5 supercell, tilted by $\theta_z$ from the $z$ axis (Fig.~\ref{Fig1}a), exhibit in-plane components $d_{\text{Ti},xy}$ with approximately equal magnitudes in each layer, while the azimuthal angle rotates progressively from one layer to the next; the out-of-plane components $d_{\text{Ti},z}$ remain largely unchanged.
This helical arrangement gives rise to structural chirality in the system. Additionally, the in-plane projection of the dipole path forms a deformed polygon rather than a perfect circle (Fig.~\ref{Fig1}b), reflecting the underlying $C_4$ symmetry of the strained crystal lattice.


These helical phases are associated with two distinct energy manifolds (green shading in Fig.~\ref{Fig1}c,d). The first, $\mathcal{M}_1$, is a lower-energy manifold corresponding to the calculated ground state. It is characterized by a dominant out-of-plane polarization and includes a quasi-continuous set of states such as $\mathcal{S}1$, $\mathcal{S}2$, and $\mathcal{S}3$, with polarization tilt angles $\theta_z$ of 4$^\circ$, 14$^\circ$, and 32$^\circ$, respectively. The second, $\mathcal{M}_2$, is a higher-energy, metastable manifold dominated by in-plane polarization, represented by states such as $\mathcal{S}4$ ($\theta_z = 87^\circ$) and $\mathcal{S}5$ ($\theta_z \approx 90^\circ$).
As the system transitions from $\mathcal{S}1$ to $\mathcal{S}5$, the out-of-plane component $d_{\text{Ti},z}$ decreases while the in-plane component $d_{\text{Ti},xy}$ increases, corresponding to a larger tilt angle $\theta_z$.
The energy landscape associated with these helical phases resembles an asymmetric double-well potential. The two basins correspond to the $\mathcal{M}_1$ and $\mathcal{M}_2$ manifolds. Within each basin, the energy variation with respect to $\theta_z$ is relatively small, allowing for a quasi-continuous evolution of the tilt angle.
The transition between the two manifolds is asymmetric: the energy barrier from $\mathcal{M}_1$ (e.g., $\mathcal{S}2$) to $\mathcal{M}_2$ (e.g., $\mathcal{S}4$) is approximately 8~meV/u.c., whereas the reverse barrier is less than 1~meV/u.c. This energy profile is consistent with a first-order structural phase transition and suggests that any field-driven switching between the two manifolds could involve notable hysteresis.


The physical origin of the two energy manifolds is examined using a phenomenological Landau-Ginzburg-Devonshire (LGD) model (see SM~\cite{SIyhh}, Sec.\MakeUppercase{\romannumeral 2}.C). The model suggests that the complex energy landscape can support multiple, nearly degenerate minima due to the high-dimensional parameter space involving coupled polarization components from both the Pb and Ti sublattices (in-plane and out-of-plane). For a fixed dipole spiral periodicity, different combinations of these sublattice polarizations can couple to produce states with comparable total energies, thereby giving rise to the $\mathcal{M}_1$ and $\mathcal{M}_2$ manifolds identified in the first-principles calculations.

\begin{figure*}
    \centering
    \includegraphics[width=1\textwidth]{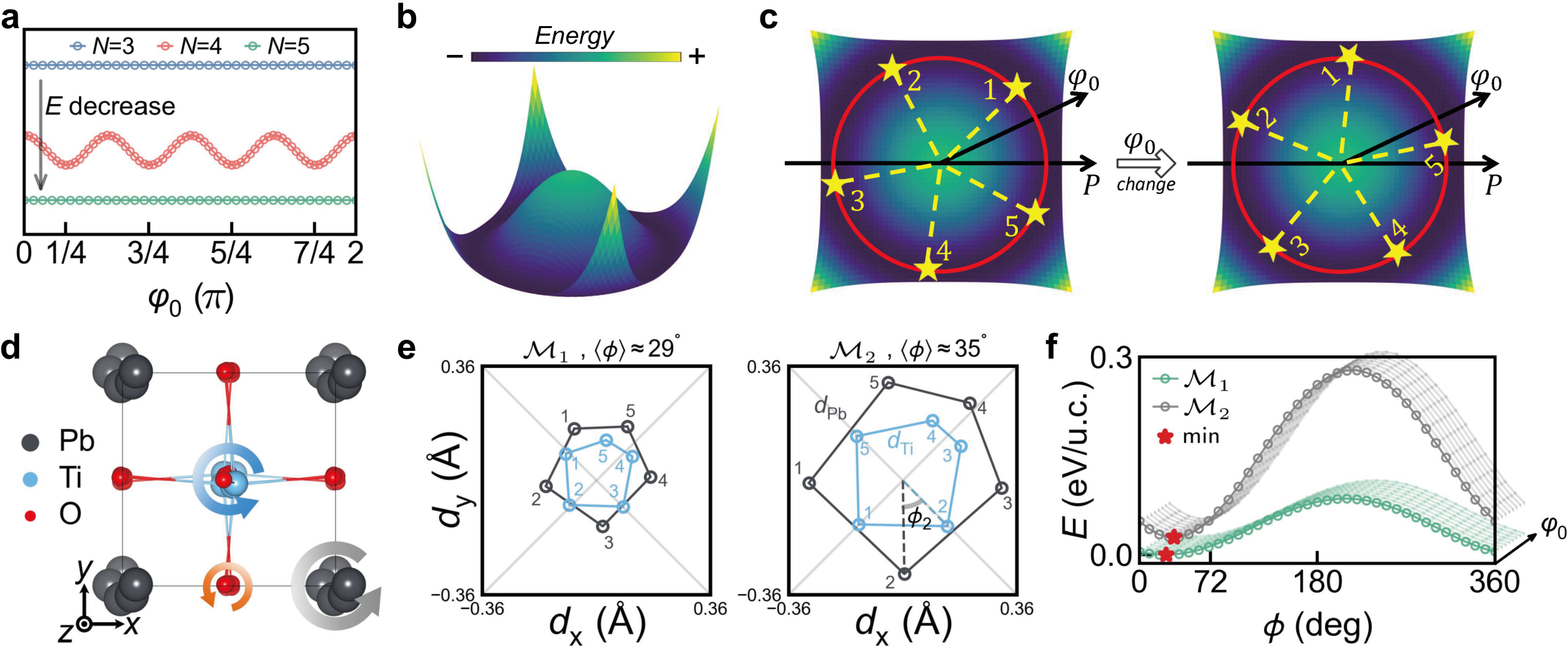}
    \caption{\textbf{Microscopic Origin and Dynamics of the Ferroelectric Double Helix.} 
    (a) Total energy as a function of the global in-plane orientation $\varphi_0$. 
    When the $\phi$ is fixed, the energy is independent of $\varphi_0$; 
    therefore, the addition of the $\varphi_0$ axis in (f) causes the curves to exhibit repetitive and consistent three-dimensional shadows.
    (b) A conceptual, three-dimensional representation of the ``Mexican-hat" potential energy surface that describes the collective rotational dynamics of an idealized spiral.
    (c) Illustration of the Goldstone-like mode for an idealized $N$=5 ``perfect pentagon" spiral. The left panel shows a schematic of this state, where the in-plane polarization vectors of the five layers (yellow stars) are equally spaced on a circle (red), with a fixed 72$^\circ$ angle between adjacent layers. The collective rotation of this idealized spiral is a zero-energy process (right panel).
    (d) Top view of the relaxed supercell, revealing the intertwined helical nature of the Pb- (gray) and Ti-sublattice (blue) polarizations, which form a double helix with a non-parallel phase angle $\langle\phi\rangle$ for both (e, left panel) $\mathcal{M}_1$ and (e, right panel) $\mathcal{M}_2$ (right) spirals. 
    (f) Total energy as a function of the relative phase angle $\phi$ between $d_{\text{Ti},xy}$ and $d_{\text{Pb},xy}$.    
    }
    \label{Fig3}
\end{figure*}

A notable feature of the dipole spiral is its ability to support a rotational zero-energy mode. The emergence of this mode arises from the general property of placing a helical distribution of vectors within a crystal lattice that exhibits a quadruple-well potential. 
For any idealized, polygonal helical arrangement (Fig.~\ref{Fig3}c, yellow stars), whether defined by the displacement of the Ti cation or by the net cell polarization, the global orientation can be specified by the in-plane polarization angle of the first layer, denoted $\varphi_0$. Once this angle is set, the orientations of all subsequent layers are determined by the helical relationship $\varphi_k = \varphi_0 + k \cdot 2\pi/N$, where $k$ is the layer index.
Although one might expect the system's total energy to depend on $\varphi_0$ due to the quadruple-well potential imposed by the strained lattice, our calculations show that for any periodicity $N \ge 3$ that is incommensurate with the underlying potential ($N \neq 4$), the total energy remains independent of $\varphi_0$ (Fig.~\ref{Fig3}a). This results in an emergent $U(1)$ symmetry and an associated rotational zero-energy mode.
The energy landscape as a function of the global angle $\varphi_0$ is analogous to a ``Mexican-hat" potential, where the brim is flat (Fig.~\ref{Fig3}b). This introduces a qualitative difference in the degrees of freedom compared to conventional ferroelectric phases. In a rhombohedral-like ($R$) or monoclinic ($M$) phase, the energy is primarily determined by the orientation of a single polarization vector, typically specified by $P_{xy}$ and $P_z$, within the potential landscape.
In contrast, the dipole spiral is characterized by the collective distribution of polarization vectors across all layers. This arrangement introduces a continuous degree of freedom: the global in-plane phase angle $\varphi_0$. The flat energy profile along this coordinate defines the rotational zero-energy mode. The corresponding excitation is a Goldstone-like mode~\cite{Goldstone62p965,Zhang21p024101}, which may contribute to an enhanced response to external fields.

A subtlety arises in the precise nature of this Goldstone-like mode. At zero temperature, the system minimizes its energy by allowing the polarization vectors to relax slightly toward the four lowest-energy wells of the $C_4$ potential. This leads to a ``deformed polygon" ground state rather than a perfectly symmetric configuration.
Although the deformation is small, it weakly breaks the emergent $U(1)$ symmetry and introduces a finite energy gap in the rotational mode. The idealized ``perfect polygon" state, which supports a true zero-energy mode, is therefore slightly higher in energy. As a result, at 0 K, the system exhibits weak pinning, and the excitation is more accurately described as a ``pseudo-Goldstone" mode.
This pinning, however, is fragile. The zero-energy character of the mode is gradually restored as the periodicity $N$ increases, causing the deformed structure to approach the idealized configuration. Moreover, thermal fluctuations ($k_B T$) can readily overcome the small pinning energy (less than 1 meV/u.c.; see SM~\cite{SIyhh}, Fig.~S6), allowing the system to explore the full rotational phase space, effectively recovering the behavior of a perfect polygon.

To better understand the microscopic nature of the dipole spiral, we performed a detailed analysis of the displacements of individual atomic sublattices. This analysis revealed a structural feature that extends beyond a simple rotation of a unit-cell dipole. We found that all three constituent sublattices, Pb, Ti, and O, exhibit helical displacement patterns along the propagation axis (Fig.~\ref{Fig3}d). This hierarchical and intertwined helical ordering of the Pb-, Ti-, and O-sublattices can be conceptualized as a one-dimensional ``self-Moir\'e" crystal. In contrast to conventional Moir\'e patterns, which are extrinsically generated at hetero-interfaces~\cite{Rafi11p12233,Cao18p43,Timmel20p166803,Bai20p1068}, this superlattice emerges intrinsically within a single, homogeneous material under uniform strain.

In perovskite oxides, coupling between adjacent layers is often described in terms of the relative rotation of oxygen octahedra, typically categorized as either in-phase or out-of-phase, with the latter generally being energetically favorable~\cite{Kim20p4944}. However, a continuous, multi-layer helical rotation of the octahedra has not been widely considered, which may have contributed to the dipole spiral being previously overlooked.
This collective chiral distortion of the lattice gives rise to two distinct local polarization vectors that form the basis of the dipole spiral: $d_{\rm Ti}$, defined by the displacement of the Ti cation relative to the center of its surrounding oxygen octahedron (BO$_6$), and $d_{\rm Pb}$, defined by the displacement within the Pb-centered AO$_{12}$ cage.
We find that these two sublattice polarizations form distinct, intertwined helices that are intrinsically non-parallel (Fig.~\ref{Fig3}e). While both $d_{\rm Pb}$ and $d_{\rm Ti}$ follow helical paths, a stable, non-zero phase angle $\langle\phi\rangle = \frac{1}{N} \sum_{k=1}^N \phi_k$ exists between them. For the $\mathcal{M}_1$ ($\mathcal{M}_2$) state, this angle is calculated to be approximately 29$^\circ$ (35$^\circ$). The non-parallelism reveals an internal, intra-cell degree of freedom and challenges the common approximation of representing unit-cell polarization as a single vector.

To quantify the energetics of this non-parallel configuration, we performed a model calculation in which an idealized ``perfect polygon'' double helix was constructed using the average $d_{\rm Pb}$ and $d_{\rm Ti}$ extracted from the fully relaxed (deformed) spiral ground state (also shown in Fig.~\ref{Fig3}a). We then computed the total energy as a function of the relative phase angle $\phi$, which corresponds to $\langle\phi\rangle$ (Fig.~\ref{Fig3}f).
The resulting energy dependence exhibits a trigonometric form, consistent with predictions from the Landau-Ginzburg-Devonshire (LGD) model (see SM~\cite{SIyhh}, Sec.~\MakeUppercase{\romannumeral 2}.D). In this double-helix configuration, while inter-helix interactions, quantified by the relative angle between the Pb and Ti helices, determine the absolute energy, they do not break the emergent $U(1)$ symmetry or gap the rotational zero-energy mode (Fig.~\ref{Fig3}f, $\varphi_0$ axis).
The calculation reveals distinct energy minima (red stars) at phase angles of approximately 27$^\circ$ and 35$^\circ$, in agreement with the values observed in the relaxed, deformed structures. This indicates that the non-parallel configuration is energetically favored. 
Furthermore, hypothetical configurations in which only a single cation sublattice forms a helix, or where both helices are aligned ($\phi = 0$), were found to be energetically unstable relative to the non-parallel ground state (see SM~\cite{SIyhh}, Fig.~S5).

To explore the microscopic origin of this energy minimum, we turned to the LGD model (see SM~\cite{SIyhh}, Sec.~\MakeUppercase{\romannumeral 2}.E). We found that the stabilization of a non-zero phase angle could only be reproduced by introducing an emergent eDMI term between the two sublattice polarizations, of the form $\propto \mathcal{D} \cdot (d_{\rm Pb} \times d_{\rm Ti})$. This suggests that the collective chiral distortion of all three atomic sublattices breaks local inversion symmetry in a way that gives rise to an intrinsic eDMI. This interaction stabilizes the non-parallel double-helix configuration and offers a microscopic explanation for its energetic preference.

\begin{figure*}
    \centering
    \includegraphics[width=1\textwidth]{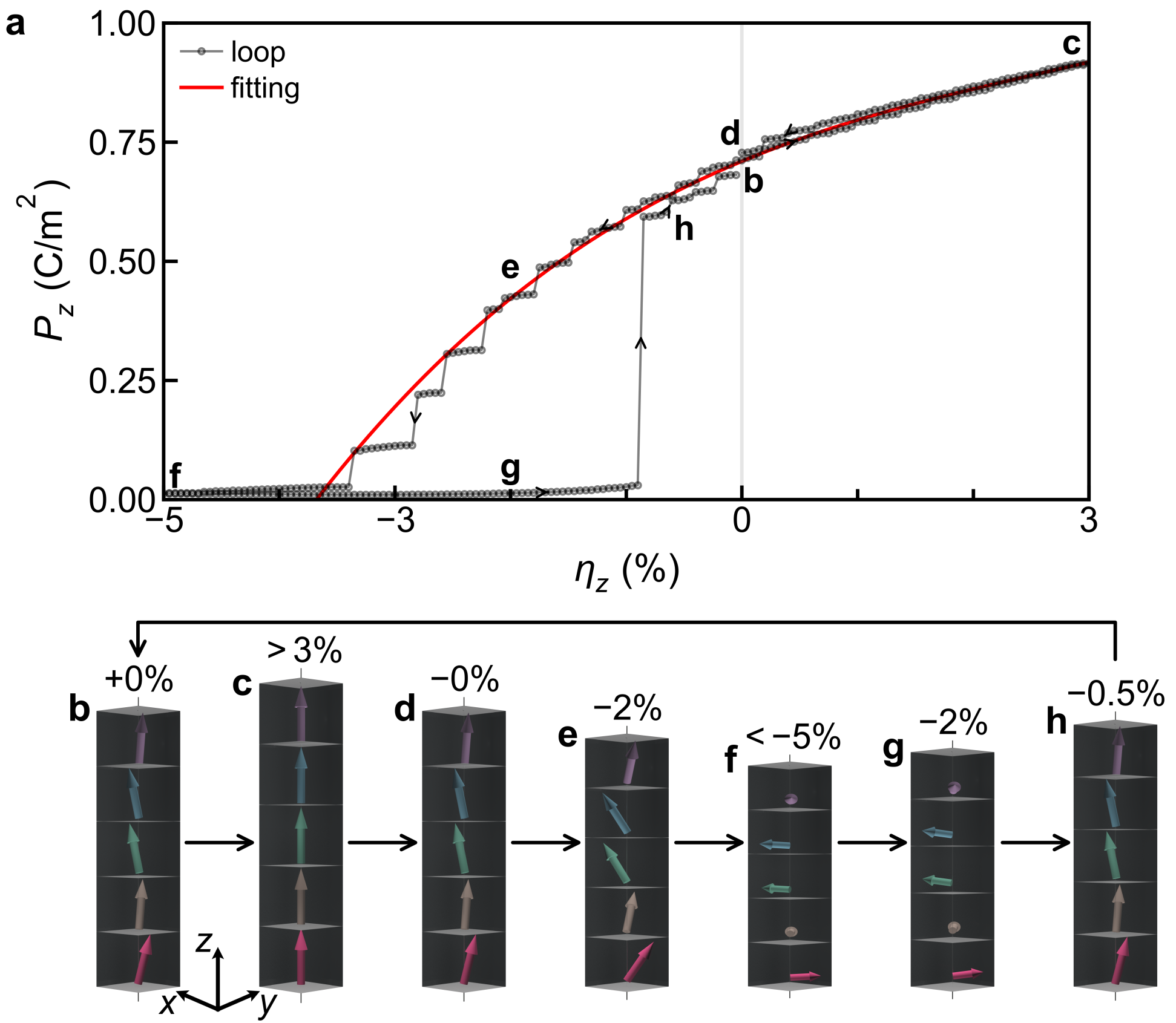}
    \caption{\textbf{Giant and Hysteretic Electromechanical Response.} (a) Calculated out-of-plane polarization ($P_z$) versus uniaxial strain ($\eta_z$) for the $\mathcal{S}$2 spiral. The reversible response near equilibrium (red curve) is highly anharmonic, yielding a giant piezoelectric coefficient $e_{33} \approx 16 \, \text{C/m}^2$. The large loop demonstrates hysteretic, first-order switching between two manifolds. (b-h) Snapshots of the dipole configurations at key points in the strain cycle, illustrating the initial $\mathcal{S}$2 state (b), the irreversible collapse to a low-$P_z$ state ($\mathcal{M}_2$) under compression (e $\to$ f), and the transformation to a topologically trivial $T$ phase under large tension (c).}
    \label{Fig2}
\end{figure*}

The unique topology of dipole spiral gives rise to an extraordinary piezoelectric response. To probe its electromechanical characteristics, we performed first-principles calculations of the out-of-plane polarization $P_z$ as a function of out-of-plane strain $\eta_z$.
The result, shown in Fig.~\ref{Fig2}a, reveals a giant and highly nonlinear piezoelectric response. Near the equilibrium state ($\eta_z$ = 0), we extract an intrinsic piezoelectric coefficient $e_{33}$ of $\approx$16 C/m$^2$, a value roughly three-fold larger than that calculated for the $T$ phase under identical strain conditions~\cite{Li96p1433,Szabo98p4321} (see SM~\cite{SIyhh}, Sec.\MakeUppercase{\romannumeral 9}).

This giant piezoelectricity originates from a unique, collective response mechanism rooted in dipole spiral's rotational pseudo-zero-energy mode. 
Unlike $R$-like or $M$ phase, which respond to an out-of-plane perturbation through a rigid and energetically costly change in their polarization orientation, the dipole spiral evolves along a much lower-energy pathway.
Instead, the pseudo-zero-energy mode provides a nearly frictionless pathway for dipole spiral to evolve. The system can simultaneously execute an in-plane rotation, which consumes negligible energy, while making an infinitesimal adjustment to its tilt angle $\theta_z$. The macroscopic piezoelectric response is thus the cumulative result of many such small, rotation-assisted tilting events, allowing the system to efficiently produce a large change in polarization for a minimal energy cost. 
This complex pathway is also responsible for the markedly anharmonic response, which directly reflects the non-parabolic shape of the flattened energy potential.

Under large cyclic strain, the system reveals its complex energy landscape through a hysteretic loop featuring first-order structural phase transitions between competing spiral manifolds $\mathcal{M}_1$ and $\mathcal{M}_2$. This hysteresis is a direct manifestation of their asymmetric energy barriers. Upon applying a critical compressive strain [(e)$\to$(f)], the high-$P_z$ ($\mathcal{M}_1$) spiral undergoes an abrupt transformation into the low-$P_z$ ($\mathcal{M}_2$) manifold. Once collapsed, the system is trapped until a critical tensile strain is applied [(g)$\to$(h)], causing it to jump back to the high-$P_z$ ($\mathcal{M}_1$) state and completing the non-volatile switching cycle (see SM~\cite{SIyhh}, Fig.~S9).
Crucially, the critical tensile strain required for this ``jump back" is not a fixed value but is dependent on the in-plane biaxial strain (see SM~\cite{SIyhh}, Fig.~S10). 
The link between helical topology and emergent functionality is proven by the system's behavior under extreme tensile strain. At $\eta_z>$3\%, 
dipole spiral reversibly transforms into $T$-like phase (Fig.~\ref{Fig2}c). In this topologically trivial phase, the piezoelectric response immediately becomes linear and its magnitude reverts to that of a standard ferroelectric. This demonstrates that the giant piezoelectricity is 
an emergent property intrinsically bound to the helical topology.

It is important to consider the influence of finite temperature on this complex electromechanical response. At 0~K, the enhanced piezoelectricity originates from the pseudo-Goldstone mode associated with the rotational degree of freedom.
At room temperature, the thermal energy ($k_{\rm B}T \approx 25$~meV) exceeds both the intra-manifold pinning potential ($\sim$1~meV) and the inter-manifold energy barrier ($\sim$8~meV). As a result, thermal fluctuations are expected to effectively restore a gapless Goldstone mode and render the $\mathcal{M}_1$ and $\mathcal{M}_2$ manifolds nearly degenerate.
This thermally-induced degeneracy permits large-amplitude fluctuations of the polarization vector, potentially spanning the full tilt angle range $\theta_z = 0^\circ$ to $90^\circ$. The collapse of the 0~K hysteresis under these conditions leads to an ultra-soft structural state, suggesting that the piezoelectric response at room temperature could be further enhanced.

\begin{figure*}
    \centering
    \includegraphics[width=1\textwidth]{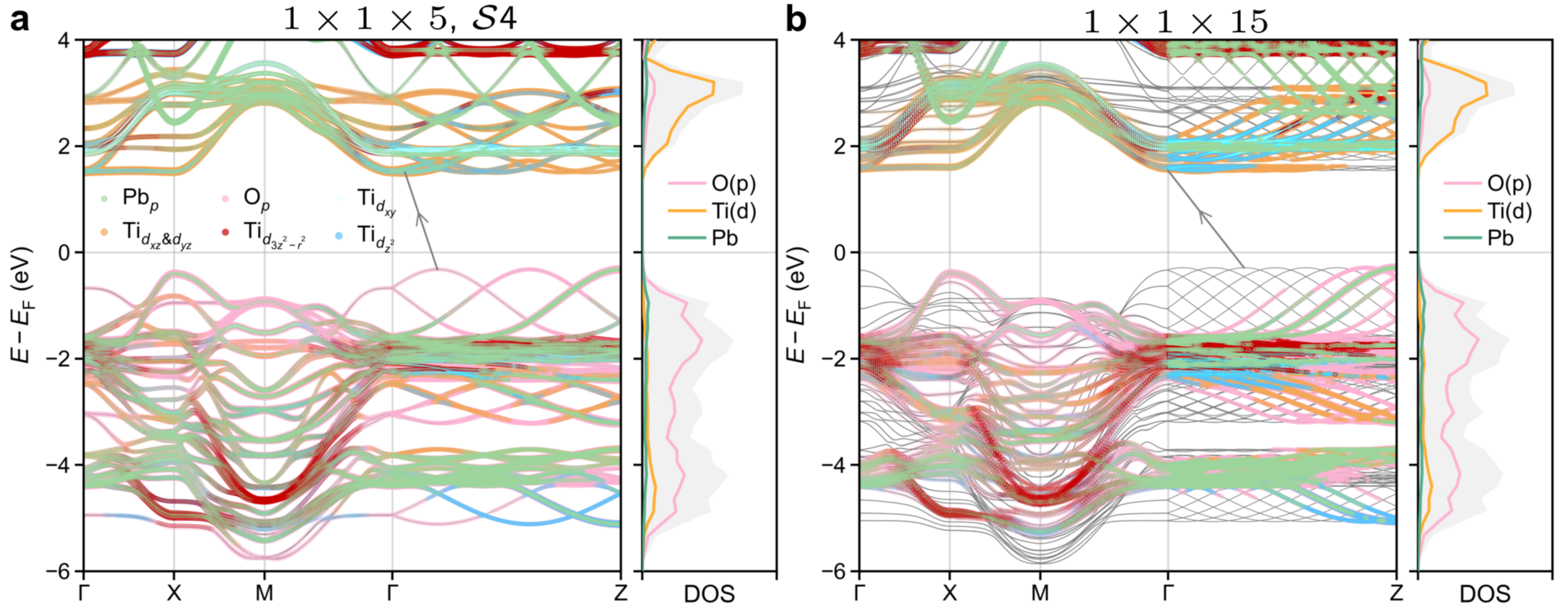}
    \caption{\textbf{Electronic Structure Reconstruction and Multi-Valley Topology.} Unfolded band structures and corresponding projected density of states (PDOS) for dipole spirals when (a) $N$=5 and (b) $N$=15. 
    A manifold of nearly-degenerate local valence band maxima (VBMs) emerges along the $\Gamma$-$Z$ direction (spiral propagation axis). 
    }
    \label{Fig4}
\end{figure*}

Beyond its electromechanical properties, the chiral dipole spiral induces a reconstruction of the material's electronic structure, giving rise to an emergent multi-valley electronic topology. Our band structure calculations, performed using a band unfolding technique~\cite{Popescu12p085201}, reveal a manifold of degenerate local valence band maxima (VBMs) along the $\Gamma$--$Z$ direction. These form a characteristic woven-shaped band structure (Fig.~\ref{Fig4}). The number of distinct VBMs, or valleys, denoted $N_v$, scales directly with the spiral periodicity $N$ via $N_v = N/2$.

It is important to distinguish this multi-valley system from a flat band, despite the visually flat dispersion observed along the short $\Gamma$--$Z$ $k$--path composed of multiple VBMs. A more definitive distinction is revealed through the density of states (DOS) analysis. In true flat-band systems, such as those observed in moir\'e heterostructures, a large number of electronic states are compressed into a narrow energy window, resulting in a sharp, high-intensity DOS peak, which plays a central role in driving strong electronic correlations~\cite{Kiesel13p126405,Van53p1189,Kang22p301,Cao18p43}.
In contrast, our calculated DOS for the dipole spiral phase decreases smoothly and monotonically toward zero at the valence band edge. This indicates that each of the $N_v$ valleys retains a normal, parabolic-like dispersion near its maximum. The total DOS is simply the superposition of these individual 3D-like valleys and therefore lacks the sharp peak associated with true electronic flatness. The key feature of this system is not a dispersionless state but rather the high degeneracy of multiple, structurally induced valleys.

This valley degeneracy could carry significant physical implications. Optically, it implies a highly degenerate excitonic ground state, which could give rise to complex fine-structure splitting and strong circular dichroism, reflecting the structure's intrinsic chirality. In terms of transport and many-body behavior, the presence of multiple valleys offers a pathway to a strongly correlated regime. Upon carrier (hole) doping, charge carriers are distributed among the $N_v$ degenerate valleys. This reduces the carrier density per valley, thereby suppressing the intra-valley kinetic energy (i.e., the Fermi energy). Meanwhile, the long-range Coulomb interaction, governed by real-space carrier separation, remains unaffected. As a result, the ratio of potential to kinetic energy is significantly enhanced, potentially driving the system into an interaction-dominated, strongly correlated limit.

A possible consequence of such an interaction-dominated regime is the formation of a Wigner crystal~\cite{Wigner34p1002}, a collective state in which charge carriers freeze into a spatial lattice to minimize their mutual Coulomb repulsion. While the multi-valley topology originates in momentum space, its correlated consequences, such as real-space charge ordering, would manifest in the spatial domain and could be experimentally detectable under appropriate conditions.
The dipole spiral thus offers a tunable platform for accessing and exploring strongly correlated electronic phenomena, enabled by its structural control over band degeneracy and carrier interactions.

In conclusion, we have discovered a chiral, non-collinear ferroelectric double helix in a strained ferroelectric oxide, characterized by intertwined, non-parallel helical polarizations of the Pb and Ti sublattices. This dipole spiral is stabilized by a collective helical lattice distortion that induces an emergent electronic Dzyaloshinskii-Moriya-like interaction, favoring a chiral ground state.
The resulting ``self-Moir\'e" structure uniquely combines robust out-of-plane ferroelectricity with transverse helical order, giving rise to two coupled functionalities: a giant piezoelectric response ($e_{33} \approx 16$~C/m$^2$) driven by a soft rotational mode, and an emergent multi-valley electronic topology at the valence band edge.
These results demonstrate a strain-driven strategy for engineering polar topology, unifying structural chirality, strong electromechanical coupling, and correlated electronic behavior in a single phase. This work opens new directions for chiral phononic materials, high-sensitivity sensors, and tunable platforms for correlated electron physics.
\\

{\em Data availability}~\textemdash~We have developed an online 
{\href{https://github.com/huiihao/Spiral_DFT}{{notebook}}}~\cite{datayhh}
on Github to share the training database, essential input and output files.
Further details about DFT calculations using \texttt{VASP} ~\cite{Kresse96p11169,Kresse96p15} can be found in Supplementary Material~\cite{SIyhh}.

\begin{acknowledgments}
We acknowledge the supports from Zhejiang Provincial Natural Science Foundation of China (LR25A040004).
The computational resource is provided by Westlake HPC Center.
\end{acknowledgments}

\bibliography{SL.bib}

\clearpage
\newpage

\clearpage
\newpage

\clearpage
\newpage

\clearpage
\newpage

\end{document}


\begin{CJK*}{UTF8}{gbsn}

\title{Supplemental Material for\\Double Helix of atomic displacements in Ferroelectric PbTiO$_3$}
\author{Yihao Hu (胡逸豪)}
\email{huyihao@westlake.edu.cn}
\affiliation{Zhejiang University, Hangzhou, Zhejiang 310058, China}
\affiliation{Department of Physics, School of Science, Westlake University, Hangzhou, Zhejiang 310030, China}

\author{Shi Liu (刘仕)}
\email{liushi@westlake.edu.cn}
\affiliation{Department of Physics, School of Science, Westlake University, Hangzhou, Zhejiang 310030, China}
\affiliation{Key Laboratory for Quantum Materials of Zhejiang Province, Department of Physics, School of Science, Westlake University, Hangzhou, Zhejiang 310024, China}
\affiliation{Institute of Natural Sciences, Westlake Institute for Advanced Study, Hangzhou, Zhejiang 310024, China}

\maketitle
\end{CJK*}

\clearpage

{
  \linespread{1.3}\selectfont   
  \tableofcontents
}

\linespread{1.36}

\clearpage
\newpage
\section{Computational Methods}
All first-principles calculations are performed with the projector augmented-wave (PAW) method~\cite{Blochl94p17953,Kresse99p1758}, using the Vienna \textit{ab initio} simulation package (\texttt{VASP})~\cite{Kresse96p11169,Kresse96p15}. The exchange-correlation functional is treated within the generalized gradient approximation of Perdew-Burke-Ernzerhof revised for solids (PBEsol) type~\cite{Perdew08p136406}.
The dipole spiral structures were modeled using 1$\times$1$\times$$N$ supercells of PbTiO$_3$, where $N$ represents the periodicity of the spiral along the [001] direction.
For a given strain state, the in-plane lattice parameters ($a$ and $b$) of a 1$\times$1$\times$$N$ supercell are fixed to simulate a specific biaxial tensile strain, while the atomic coordinates and out-of-plane lattice constant are fully optimized. 
This setup closely resembles the application of orthogonal strains to freestanding membranes, which is a common  scenario  in experimental settings~\cite{Xu20p3141,Hong20p71,Han23p2808}.
To access competing spiral states, multiple initial configurations with polarization pointing with different $\theta_z$ are used.
A kinetic energy cutoff of 800 eV, a $k$-point spacing of 0.3 \AA$^{-1}$ for the Brillouin zone integration, and a force convergence threshold of 0.001 eV/\AA~are used to ensure the convergence of energy and atomic forces.

The piezoelectric coefficient $e_{33}$ was calculated by numerically differentiating the out-of-plane polarization $P_z$ with respect to the out-of-plane strain $\eta_z$ ($e_{33}$ = $\partial P_z / \partial \eta_z$). At each $\eta_z$ step, the atomic positions were fully relaxed while keeping all lattice parameters fixed. 

The polarization of each unit cell is estimated using the following formula, 
\[\mathbf{p}^m=\frac{1}{V_{\rm uc}}\left[\frac{1}{8} \mathbf{Z}_{\mathrm{Pb}}^* \sum_{k=1}^8 \mathbf{r}_{\mathrm{Pb}, k}^m+\mathbf{Z}_{\mathrm{Ti}}^* \mathbf{r}_{\mathrm{Ti}}^m+\frac{1}{2} \mathbf{Z}_{\mathrm{O}}^* \sum_{k=1}^6 \mathbf{r}_{\mathrm{O}, k}^m\right]\]
where $\mathbf{p}^m$ is the polarization of unit cell $m$, $V_{\rm uc}$ is the volume of the unit cell, $\mathbf{Z}_{\mathrm{Pb}}^*, \mathbf{Z}_{\mathrm{Ti}}^*$, and $\mathbf{Z}_{\mathrm{O}}^*$ are the Born effective charges of Pb, Ti and O atoms, $\mathbf{r}_{\mathrm{Pb}, k}^m, \mathbf{r}_{\mathrm{Ti}}^m$, and $\mathbf{r}_{\mathrm{O}, k}^m$ are the instantaneous atomic positions in unit cell $m$ from DFT calculations. Here, the local polarization $\mathbf{p}^m$ is defined as the local electric dipole divided by $V_{\rm uc}$.

Electronic band structures were calculated along high-symmetry paths in the Brillouin zone, and a band unfolding technique~\cite{Popescu12p085201} was employed to project the supercell bands back onto the primitive cell's Brillouin zone for clear interpretation.

\clearpage
\newpage
\section{Landau-Ginzburg-Devonshire (LGD) Model for the Double-Helix Spiral}
\label{SecLGD}
To provide a deeper physical insight into the stability of the non-parallel double-helix structure revealed by our first-principles calculations, we developed a Landau-Ginzburg-Devonshire (LGD) type phenomenological model. This model considers the total free energy of the system as a sum of the self-energy of individual Pb and Ti sublattice helices and the interaction energy between them.

\subsection{Free Energy of a Single Helical Chain}
Following the framework established for ``single-chain" spiral~\cite{Hu24p046802}, the average free energy per layer for a single helical chain (denoted by subscript $l$, where $l=\text{Pb}$ or Ti) with periodicity $N$ can be expressed as:
\begin{equation}
    \langle f_l \rangle = \langle f_{l,\text{loc}} \rangle + \langle f_{l,\text{grad}} \rangle
\end{equation}
The local energy contribution, $\langle f_{l,\text{loc}} \rangle$, which depends on the in-plane azimuthal orientation $\phi_l^0$ of the first layer, averages to:
\begin{equation}
    \langle f_{l,\text{loc}} \rangle = 
    \begin{cases}
       \mathcal{A}_l\cos(4\phi_l^0) +\mathcal{B}_l, & N=1,2,\text{or } 4,\\
       \mathcal{B}_l, & N=3,\text{or} >4.
    \end{cases}
\end{equation}
where $\mathcal{A}_l$ represents the in-plane anisotropy energy and $\mathcal{B}_l$ is an isotropic energy term dependent on the polarization components ($p_{l,xy}, p_{l,z}$) and Landau coefficients ($\alpha_{l,ij...}$). 
\begin{equation}
\begin{aligned}
\mathcal{A}_l=&~\frac{1}{8}p_{l,xy}^4(2\alpha_{l,11}-\alpha_{l,12})+ \frac{1}{8}p_{l,xy}^6(3\alpha_{l,111}-\alpha_{l,112})+ \frac{1}{8}p_{l,xy}^4 p_{l,z}^2(2\alpha_{l,112}-\alpha_{l,123})\\
\mathcal{B}_l=&~\alpha_{l,1} p_{l,xy}^2+\alpha_{l,1} p_{l,z}^2+\alpha_{l,12}p_{l,xy}^2p_{l,z}^2+\alpha_{l,11}p_{l,z}^4+\alpha_{l,111}p_{l,z}^6+\alpha_{l,112}p_{l,xy}^2p_{l,z}^4\\
&~+\frac{1}{8}p_{l,xy}^4(6\alpha_{l,11}+\alpha_{l,12})+\frac{1}{8}p_{l,xy}^6(5\alpha_{l,111}+\alpha_{l,112})+\frac{1}{8}p_{l,xy}^4 p_{l,z}^2(6\alpha_{l,112}+\alpha_{l,123}) \\
\end{aligned}
\end{equation}
The term $\mathcal{A}_l\cos(4\phi_l^0)$ reflects the four-fold symmetry of the underlying lattice. For $N>4$ and $N=3$, this anisotropic term averages to zero over the whole spiral, leading to a rotationally invariant energy landscape. The gradient energy, arising from the polarization discontinuity between adjacent layers, is given by $\langle f_{l,\text{grad}} \rangle = \mathcal{C}_l\sin^2(\pi/N)$, where $\mathcal{C}_l = 4g_l(p_{l,xy})^2$ and $g_l$ is the gradient energy coefficient. 

Thus, for our primary interest where $N \ge 3$ and $N \ne 4$, the total self-energy of the two non-interacting chains is:
\begin{equation}
    \langle f_{\text{self}} \rangle = \sum_{l=\text{Pb,Ti}} \langle f_l \rangle = (\mathcal{B}_{\text{Pb}}+\mathcal{B}_{\text{Ti}}) + (\mathcal{C}_{\text{Pb}}+\mathcal{C}_{\text{Ti}})\sin^2\left(\frac{\pi}{N}\right)
\end{equation}

\subsection{Interaction Energy of the Double Helix}
The interaction between the A- and B-chain dipoles primarily originates from a complex superposition of two contributions.
\begin{itemize}
    \item \textbf{Electrostatic Dipole-Dipole Interaction}: This is the classical, intuitive interaction whose effect is strongly dependent on the relative position and orientation of the two dipoles. It is characterized as being strongly anisotropic and long-range (decaying as $1/r^3$). The sign of the interaction energy depends on the alignment; a ``head-to-tail'' arrangement results in a negative energy (attractive, ferroelectric coupling), while a ``side-by-side'' arrangement can be positive or negative depending on the specific angle.

    \item \textbf{Short-Range Covalent/Quantum Mechanical Effect}: This effect stems from the wavefunction overlap and hybridization of adjacent atomic orbitals. In perovskites, the covalent bond formation between Ti d-orbitals and O p-orbitals is a core source of ferroelectricity. The A-site ion (e.g., Pb 6s orbitals) can also hybridize with the oxygen octahedron. This interaction is short-range but can be very strong, typically favoring a synergistic, coherent motion of atoms that optimizes the covalent bond network. Therefore, this mechanism strongly promotes a ferroelectric alignment, contributing a negative value to the coupling coefficients.
\end{itemize}

The crucial physics of the double helix is captured by the interaction energy between the Pb and Ti spirals. We propose this interaction consists of two main contributions: a symmetric exchange-like coupling and an antisymmetric Dzyaloshinskii-Moriya-like (eDMI) coupling.

\subsubsection{Symmetric Exchange-like Coupling}
This term represents all interactions that favor a collinear (parallel or antiparallel) alignment of the sublattice polarizations, primarily driven by short-range covalent bonding and long-range electrostatic interactions. We model this with a Heisenberg-like form, including both intra-layer (quasi-local, coefficient $\xi_d$) and inter-layer (gradient-like, coefficient $\lambda_d$) contributions.

The quasi-local term describes the coupling between dipoles $\boldsymbol{p}_A^k$ and $\boldsymbol{p}_B^k$ within the same layer index $k$. The energy density for layer $k$ is given by:
\begin{equation}
    f_{\text{inter,loc}}^{k} = \boldsymbol{p}_A^k \cdot \boldsymbol{\xi} \cdot \boldsymbol{p}_B^k
\end{equation}
where $\boldsymbol{\xi}$ is a $3 \times 3$ coupling tensor. This can be decomposed into out-of-plane ($z$) and in-plane ($xy$) components.

The $z$-component is:
\begin{equation}
    f_{\text{inter,loc}}^{k}|_{z} = \xi_z p_{A,z}^k p_{B,z}^k
\end{equation}
The $xy$-component, including isotropic ($\xi_d$) and anisotropic ($\xi_\alpha$) parts, is:
\begin{equation}
\begin{aligned}
f_{\text{inter,loc}}^{k}|_{xy} = & \xi_d (p_{A,x}^k p_{B,x}^k + p_{A,y}^k p_{B,y}^k) + \xi_\alpha (p_{A,x}^k p_{B,y}^k + p_{A,y}^k p_{B,x}^k) \\
   = & \xi_d p_{A,xy} p_{B,xy} \cos(\Delta\phi_0) + \xi_\alpha p_{A,xy} p_{B,xy} \sin(\Sigma \phi_0+2k\delta)
\end{aligned}
\end{equation}
where we define $\Delta \phi_0 (=\phi ~\text{in the main text})= \phi_{A}^{0}-\phi_{B}^{0}$ and $\Sigma \phi_0 = \phi_{A}^{0}+\phi_{B}^{0}$.
The total local interaction energy for layer $k$ is the sum of these components:
\begin{equation}
f_{\text{inter,loc}}^{k} = \left[\xi_d \cos(\Delta\phi_0) + \xi_\alpha \sin(\Sigma \phi_0+2k\delta)\right]p_{A,xy} p_{B,xy} + \xi_z p_{A,z} p_{B,z}
\end{equation}
Averaging over the entire supercell, the oscillating anisotropic term vanishes, yielding the average local interaction energy per layer :
\begin{equation}
\langle f_{\text{inter,loc}} \rangle = \frac{1}{N}\sum _{k=1}^N  f_{\text{inter,loc}}^{k} = \xi_d \cos(\Delta\phi_0) p_{A,xy} p_{B,xy} + \xi_z p_{A,z} p_{B,z}
\end{equation}
During the derivation, the identity, $\sum_{k=1}^N\sin{(\Sigma\phi_0+2k\delta)}=0$, is used, see proof in \texttt{APPENDIX}.
The quasi-local coupling coefficient $\xi_d$ describes the interaction between $\boldsymbol{p}_A^k$ and $\boldsymbol{p}_B^k$ within the same unit cell index $k$. This represents the nearest-neighbor interaction between the A and B chains, such as the interaction between the body-centered B-site ion and the surrounding corner A-site ions in a perovskite. In most ferroelectric perovskites like PbTiO$_3$, PZT, and BaTiO$_3$, the A-site and B-site cation displacements are collinear and contribute synergistically to the total polarization. This behavior is dominated by short-range orbital hybridization effects. Consequently, $\xi_d$ is expected to be negative ($\xi_d < 0$), indicating a ferroelectric coupling that favors parallel alignment of the local dipoles.

This term describes the coupling between dipoles in adjacent layers, i.e., between layer $k$ and $k+1$. The total energy for the $k$-th interaction slice is the sum of $\boldsymbol{p}_A^k \leftrightarrow \boldsymbol{p}_B^{k+1}$ and $\boldsymbol{p}_A^{k+1} \leftrightarrow \boldsymbol{p}_B^k$ couplings.

The $z$-component is:
\begin{equation}
\begin{aligned}
f_{inter,grad}^{k}|_{z} = & \lambda_z (p_{A,z}^k p_{B,z}^{k+1} + p_{A,z}^{k+1} p_{B,z}^k)\\
   = & \lambda_z (p_{A,z} p_{B,z} + p_{A,z} p_{B,z}) \\
   = & 2\lambda_z p_{A,z} p_{B,z}
\end{aligned}
\end{equation}

The $xy$-component, including isotropic ($\lambda_d$) and anisotropic ($\lambda_\alpha$) parts.

For the polarization at the $k$-th layer of chain A and the polarization at the $(k+1)$-th layer of chain B:
\begin{equation}
\begin{aligned}
\text{isotropic:~~} & \lambda_d (p_{A,x}^{k} p_{B,x}^{k+1} + p_{A,y}^{k} p_{B,y}^{k+1})\\
   &  =  \lambda_d p_{A,xy} p_{B,xy} \cos{(\phi_{A}^{k}-\phi_{B}^{k+1})} \\
\text{anisotropic:~~} & \lambda_\alpha (p_{A,x}^{k} p_{B,y}^{k+1} + p_{A,y}^{k} p_{B,x}^{k+1}) \\
 &   = \lambda_\alpha [p_{A,xy}\cos{\phi_{A}^{k}} p_{B,xy}\sin{\phi_{B}^{k+1}} + p_{A,xy}\sin{\phi_{A}^{k}} p_{B,xy}\cos{\phi_{B}^{k+1}} ] \\
 &   = \lambda_\alpha p_{A,xy} p_{B,xy} \sin{(\phi_{A}^{k}+\phi_{B}^{k+1})}
\end{aligned}
\end{equation}
where $\phi_{l}^{k} = \phi_{l}^{0}+k\delta = \phi_{l}^{0}+k\frac{2\pi}{N}$ ($l$=A,B).
Similarly, for the polarization at the $(k+1)$-th layer of chain A and the $k$-th layer of chain B, the same reasoning applies:
\begin{equation}
\begin{aligned}
\text{isotropic:~~} & \lambda_d (p_{A,x}^{k+1} p_{B,x}^{k} + p_{A,y}^{k+1} p_{B,y}^{k})\\
  &   =  \lambda_d p_{A,xy} p_{B,xy} \cos{(\phi_{A}^{k+1}-\phi_{B}^{k})} \\
\text{anisotropic:~~} & \lambda_\alpha (p_{A,x}^{k+1} p_{B,y}^{k} + p_{A,y}^{k+1} p_{B,x}^{k}) \\
 &   = \lambda_\alpha p_{A,xy} p_{B,xy} \sin{(\phi_{A}^{k+1}+\phi_{B}^{k})}
\end{aligned}
\end{equation}

Sum the isotropic and anisotropic contributions, we arrive at the total gradient interaction energy for the $k$-th slice:
\begin{equation}
f_{\text{inter,grad}}^k = 2 \left[\lambda_d \cos(\Delta\phi_0)\cos(\delta) + \lambda_\alpha \sin(\Sigma\phi_{0}+(2k+1)\delta) \right] p_{A,xy}p_{B,xy} + 2\lambda_z p_{A,z}p_{B,z}
\end{equation}
Averaging this expression over the supercell, the oscillating anisotropic term again vanishes, resulting in the average gradient interaction energy per layer:
\begin{equation}
\langle f_{\text{inter,grad}} \rangle = \frac{1}{N}\sum _{k=1}^N  f_{\text{inter,grad}}^{k} = 2 \lambda_d \cos(\Delta\phi_0)\cos(\delta) p_{A,xy}p_{B,xy} + 2\lambda_z p_{A,z}p_{B,z}
\end{equation}
The gradient-like coupling coefficient $\lambda_d$ describes the next-nearest neighbor interaction between the A and B chains (e.g., between $\boldsymbol{p}_A^k$ and $\boldsymbol{p}_B^{k+1}$). For this inter-layer coupling, short-range covalent effects are weaker, while long-range electrostatic contributions become relatively more important. However, to establish long-range ferroelectric order, this coupling must also be cohesive. To ensure a smooth spatial distribution of polarization (for either a uniform or helical state), the inter-layer coupling should also be ferroelectric. It is therefore reasonable to assume that $\lambda_d$ is typically negative ($\lambda_d < 0$). However, its strength is generally expected to be weaker than the quasi-local coupling, i.e., $|\lambda_d| < |\xi_d|$.

Combining the average local and gradient-like terms, the average interaction energy per layer is:
\begin{equation}
    \langle f_{\text{inter}} \rangle = (\xi_d + 2\lambda_d \cos\delta) p_{\text{Pb},xy} p_{\text{Ti},xy} \cos(\Delta\phi_0) + z\text{-term}
    \label{eq:inter}
\end{equation}
where $\Delta\phi_0 = \phi_{\text{Pb}}^0 - \phi_{\text{Ti}}^0$ is the relative phase angle between the two spirals, and $\delta = 2\pi/N$. Based on the strong ferroelectric coupling in PbTiO$_3$, both $\xi_d$ and $\lambda_d$ are expected to be negative, favoring a parallel alignment ($\Delta\phi_0=0$).

\subsubsection{Antisymmetric eDMI Coupling}
In addition to the symmetric exchange that favors collinear alignment, an antisymmetric coupling, analogous to the Dzyaloshinskii-Moriya interaction (DMI) in magnetism, can arise in systems with broken inversion symmetry. This term energetically favors a non-parallel (e.g., perpendicular) alignment of the dipoles.

The general form of the eDMI energy is $\boldsymbol{\mathcal{D}} \cdot (\boldsymbol{p}_A \times \boldsymbol{p}_B)$. The direction of the eDMI vector $\boldsymbol{\mathcal{D}}$ is not arbitrary but is strictly constrained by the crystal's symmetry. For the helical structure discussed, we primarily consider only the $z$-component, $\mathcal{D}_z$, for two main reasons:

\begin{itemize}
    \item \textbf{Symmetry Constraints}: The constraints imposed by crystal symmetry on the direction of the eDMI vector $\boldsymbol{\mathcal{D}}$ are described by Moriya's rules. For a system with a high-symmetry $z$-axis, the in-plane components of the eDMI vector are often forbidden. For instance, if the $z$-axis is a four-fold rotation axis, a 90$^\circ$ rotation transforms $\mathcal{D}_x \rightarrow \mathcal{D}_y$ and $\mathcal{D}_y \rightarrow -\mathcal{D}_x$. The only solution that simultaneously satisfies $\mathcal{D}_x = \mathcal{D}_y$ and $\mathcal{D}_y = -\mathcal{D}_x$ is $\mathcal{D}_x = \mathcal{D}_y = 0$. 
    \begin{equation}
    \underbrace{ \mathcal{D}_x\xrightarrow[\circlearrowright]{90^\circ}}_{\mathcal{D}_x=\mathcal{D}_y} \mathcal{D}_y \underbrace{
    \xrightarrow[\circlearrowright]{90^\circ}-\mathcal{D}_x}_{\mathcal{D}_y=-\mathcal{D}_x}
    \end{equation}
    Furthermore, if the $xy$-plane is a mirror plane, the $\boldsymbol{\mathcal{D}}$ vector must be perpendicular to this plane, i.e., pointing in the $z$-direction, resulting in $\boldsymbol{\mathcal{D}}=(0,0,\mathcal{D}_z)$. From a fundamental symmetry perspective, for a chiral structure propagating along the $z$-axis, the crystal symmetry often filters out the x and y components of the eDMI, leaving only $\mathcal{D}_z$.

    \item \textbf{Geometric Constraints}: Even without considering the symmetry constraints on the $\boldsymbol{\mathcal{D}}$ vector, the geometry of the helical structure itself makes the $z$-component of the eDMI the most significant part. The polarization vectors are primarily rotating in the $xy$-plane. According to the right-hand rule, the cross product of two vectors rotating in the $xy$-plane naturally points in the $z$-direction, perpendicular to the plane. From a phenomenological geometry perspective, for a spiral twisting in the $xy$-plane, the most significant contribution of its cross product is also in the $z$-direction, while the in-plane components average to zero over a full period.
\end{itemize}

Based on the reasons above, the eDMI interaction energy for the $k$-th layer, $f_{DMI}^{k}$, is simplified to its $z$-component:
\begin{align*}
    f_{DMI}^{k} &= \mathcal{D}_{z} \cdot \left(\boldsymbol{p}_{A}^{k} \times \boldsymbol{p}_{B}^{k}\right)_{z} \\
    &= \mathcal{D}_{z} (p_{A,x}^k p_{B,y}^k - p_{A,y}^k p_{B,x}^k) \\
    &= \mathcal{D}_{z} p_{A,xy} p_{B,xy} \sin(\phi_{B}^{k}-\phi_{A}^{k}) \\
    &= -\mathcal{D}_{z} p_{A,xy} p_{B,xy} \sin(\phi_{A}^{k}-\phi_{B}^{k}) \\
    &= -\mathcal{D}_{z} p_{A,xy} p_{B,xy} \sin(\Delta \phi_{0})
\end{align*}
Since the resulting expression depends only on the initial phase difference $\Delta\phi_0$, which is constant for all layers, the energy contribution is the same for every layer. Therefore, the average energy per layer is equal to the energy of a single layer:
\begin{equation}
\langle f_{DMI} \rangle = f_{DMI}^{k} = -\mathcal{D}_{z} p_{A,xy} p_{B,xy} \sin(\Delta \phi_{0})
\end{equation}


\subsection{Total Free Energy of the Double-Chain System}
The total average free energy per layer for the double-chain system is obtained by summing the self-energy of the A and B chains ($\langle f_{\text{self}} \rangle$), the symmetric inter-chain interaction energy ($\langle f_{\text{inter}} \rangle$), and the antisymmetric eDMI energy ($\langle f_{\text{DMI}} \rangle$):
\begin{equation}
    \langle f_{\text{total}} \rangle = \langle f_{\text{self}} \rangle + \langle f_{\text{inter}} \rangle + \langle f_{\text{DMI}} \rangle
\end{equation}

\subsection{Optimization of the Relative Phase Angle}
To find the optimal relative phase angle $\Delta\phi_0$ that minimizes the free energy, we can isolate the terms in $\langle f_{\text{total}} \rangle$ that depend on it. These terms arise from the inter-chain couplings:
\begin{equation}
    \langle f_{\text{total}}(\Delta\phi_0) \rangle = \left[ (\xi_d + 2\lambda_d \cos\delta) \cos(\Delta\phi_0) - \mathcal{D}_{z} \sin(\Delta \phi_{0}) \right] p_{A,xy}p_{B,xy} + \text{const.}
    \label{eq:angle_dependence}
\end{equation}
\textbf{The trigonometric form of this energy dependence is consistent with curves obtained from DFT calculations.} We can find the optimal angle by differentiating this expression with respect to $\Delta\phi_0$ and setting the result to zero:
\begin{equation}
    \frac{d\langle f_{\text{total}}(\Delta\phi_0) \rangle}{d(\Delta\phi_0)} = 0
\end{equation}
\begin{equation}
    \implies \left[ -(\xi_d + 2\lambda_d \cos\delta) \sin(\Delta\phi_0) - \mathcal{D}_{z} \cos(\Delta \phi_{0}) \right] p_{A,xy}p_{B,xy}  = 0
\end{equation}
This equation has two types of solutions. If $p_{A,xy}p_{B,xy}=0$, the system lacks a helical component and is in a conventional ferroelectric or paraelectric phase. For a helical structure where $p_{A,xy}p_{B,xy}\neq 0$, we can solve for the optimal angle $\Delta\phi_0|_{\text{opt}}$:
\begin{gather}
    (\xi_d + 2\lambda_d \cos\delta)\sin(\Delta\phi_0) + \mathcal{D}_{z} \cos(\Delta \phi_{0}) = 0 \\
    \implies\tan(\Delta\phi_0|_{\text{opt}}) = -\frac{\mathcal{D}_z}{\xi_d+2\lambda_d \cos\delta}
    \label{eq:optimal_angle}
\end{gather}
Here, the term $(\xi_d + 2\lambda_d \cos\delta)$ can be interpreted as the effective symmetric coupling (Heisenberg-like), which represents the sum of all interactions favoring a collinear (parallel or antiparallel) alignment of the dipoles. The term $\mathcal{D}_{z}$ is the effective antisymmetric coupling (eDMI), representing all interactions that favor a perpendicular alignment.
This key result elegantly demonstrates that the equilibrium phase angle is determined by the ratio of the effective antisymmetric eDMI coupling ($\mathcal{D}_z$) to the effective symmetric exchange-like coupling ($\xi_d + 2\lambda_d \cos\delta$). It perfectly explains why a stable, non-parallel state with a specific, non-zero phase angle is the ground state of the system, providing a direct link between the microscopic interactions and the observed topological structure.

\subsection{Analysis of the Optimal Angle}
The optimal angle is determined by the competition between these symmetric and antisymmetric couplings. The effective symmetric coupling itself depends on the spiral wavelength $N$ through the term $\delta = 2\pi/N$.

For instance, when $N=3$, we have $\cos\delta = -1/2$, so the effective coupling becomes $\xi_d - \lambda_d$. Given that typically $|\lambda_d| < |\xi_d|$, this expression is expected to be negative for a ferroelectric system. For cases where $N \ge 3$, the term $(\xi_d+2\lambda_d \cos\delta)$ is generally negative. As $N$ increases ($\delta\to0$), the absolute value of this symmetric coupling term, $|\xi_d+2\lambda_d \cos\delta|$, monotonically increases. Consequently, the ratio $|\mathcal{D}_z / (\xi_d+2\lambda_d \cos\delta)|$ decreases, leading to a monotonic decrease in the optimal angle $|\Delta\phi_0|_{\text{opt}}|$. However, as long as $\mathcal{D}_z \neq 0$, the angle remains non-zero.

\subsection{Emergent Collectivity in a Three-Polarization System}
This situation highlights the emergence of a collective phenomenon, the rotational zero-energy mode, from a minimal system of just three interacting polarizations, a concept analogous to how collective behavior can arise in other few-body systems, such as those of three electrons~\cite{Shaju25p928}.

\clearpage
\newpage
\section{Winding Number and dipole spiral's topological invariant}
For a system where the polarization is confined to a plane ($\mathbb{S}^1$, a one-dimensional sphere or circle), the topological invariant can be described by the Winding Number $\omega$. It is formally defined as a line integral along a closed loop $L$:
\begin{equation}
    \omega=\frac{1}{2 \pi} \int_{L} \frac{d \theta}{d \mathbf{l}} \cdot d \mathbf{l}
\end{equation}
where $\theta$ represents the angle of the order parameter with respect to the $x$-axis. It precisely quantifies the number of times a vector field (such as the polarization vector in ferroelectric materials or the magnetic moment in magnetic materials) winds around a central point (the vortex core).

To fully capture the unique topology of the conical spiral, we introduce a new topological invariant, the Helical-Ferroelectric Invariant (HFI), defined as $\chi = (W, \sigma_z)$. Here, $W$ is the conventional winding number that quantifies the chirality of the in-plane polarization helix ($W=\pm1$)~\cite{Mermin79p591}, while $\sigma_z = sign(P_z)$ is an axial polarization order parameter that describes the system's out-of-plane ferroelectricity ($\sigma_z=\pm1$). 

\begin{figure}[htbp]
\centering
\includegraphics[width=1\textwidth]{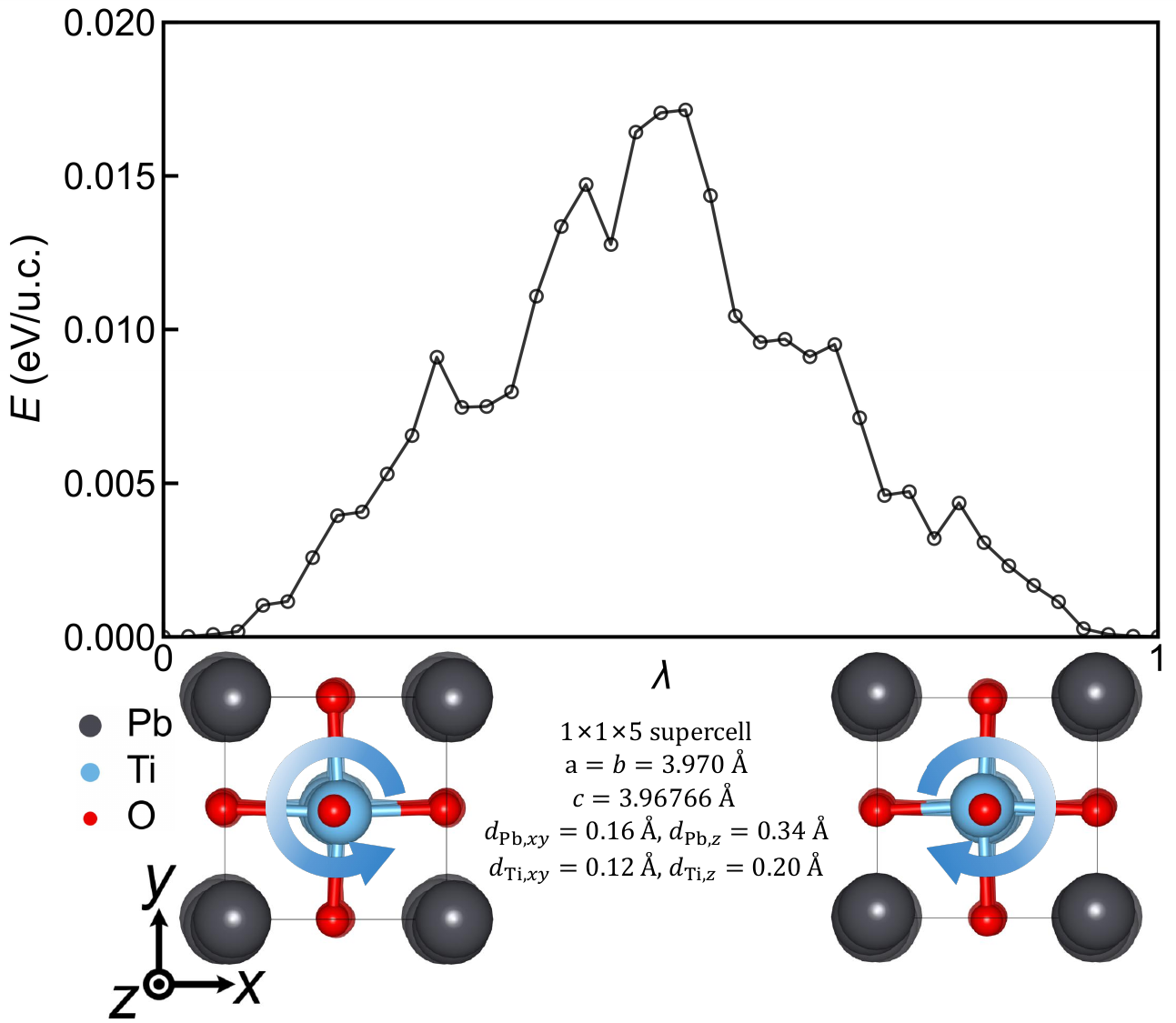}
\caption{Energy landscape along a linear interpolation path between anti-clockwise dipole spiral and clockwise dipole spiral. The insets are the top view of the $1\times1\times5$ PbTiO$_3$ ($a=b=3.970$ \AA).
}
\label{FigS3.5}
\end{figure}

Unlike the winding number alone, which only describes the planar topology, the HFI $\chi$ completely characterizes the symmetry of the conical structure by encoding both its chirality and polarity. This new invariant distinguishes four distinct, topologically protected ground states $(\pm1, \pm1)$ that cannot be transformed into one another via continuous deformation. 
The manifold $\mathcal{M}_1$ and $\mathcal{M}_2$ belongs to the $(+1,+1)$ category.
The energy barrier between the dipole spiral of the $(+1, +1)$ category and that of the $(-1, +1)$ category is quite high when out-of-plane lattice parameter is fixed. ($\sim$17 meV/u.c., Fig.~\ref{FigS3.5}).
The HFI thus provides a more complete classification scheme for this new family of polar topologies and establishes a framework for exploring potential multi-state memory applications.

\clearpage
\newpage


\section{Additional DFT modeling of dipole spirals}
\begin{figure}[htbp]
\centering
\includegraphics[width=1\textwidth]{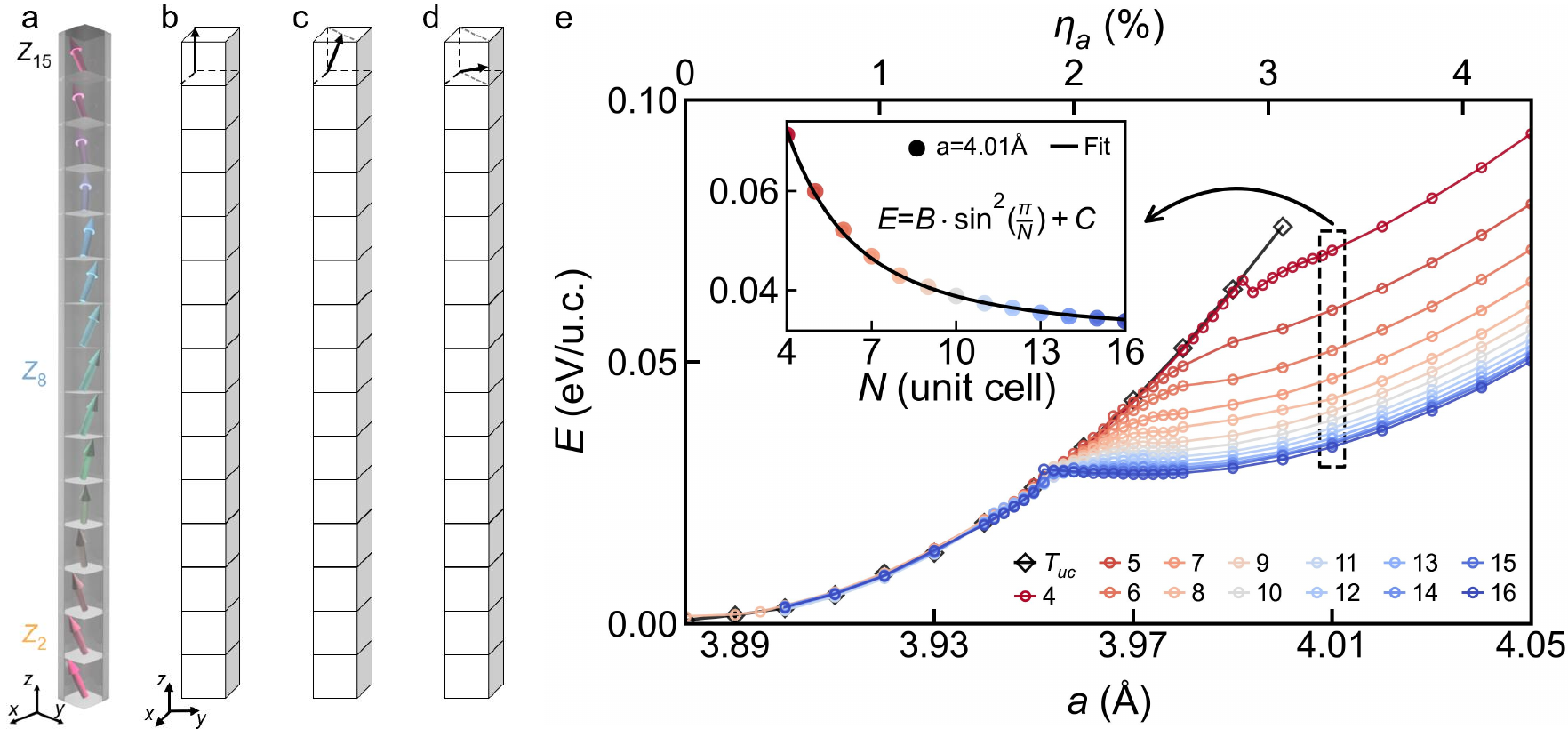}
\caption{Schematics of a \textbf{(a)} dipole spiral, \textbf{(b)} singe-domain [001], \textbf{(c)} $M_A$ and \textbf{(d)} $M_B$ states modeled with  $1\times1\times15$ supercells in DFT. \textbf{(e)} Total energy versus in-plane lattice parameter $a$, showing that spiral states (colored) become the ground state over the conventional $T$ phase (black) for $a > 3.96 \, \text{\AA}$. The inset confirms the energy dependence on periodicity $N$ is perfectly described by an LGD model.}
\label{FigS0}
\end{figure}

In \texttt{Supplemental Material} of our previous work [\href{https://journals.aps.org/prl/abstract/10.1103/PhysRevLett.133.046802}{PRL \textbf{133}, 046802 (2024)}], we have already demonstrated that the dipole spiral is the ground state with the globally lowest energy under certain conditions.
We compute the DFT energies of  $1\times1\times15$ supercells in four different polar states: (1) a dipole spiral with dipoles rotating progressively around the $z$-axis, (2) a single-domain [001] state with all unit cells having polarization aligned along [001], (3) a single-domain $M_A$ $[uu1]$ state, and (4) a single-domain $M_B$ $[11u]$ state, as depicted in Fig.~\ref{FigS0}. The in-plane lattice constants are fixed at $a_{\mathrm{IP}}= b_{\mathrm{IP}}=3.948$~\AA, while the out-of-plane lattice constant and atomic positions aer fully relaxed.
As shown in Table~\ref{DFTenergy}, the dipole spiral is lower in energy compared to the other three single-domain states, further corroborating results from MD simulations. It is noteworthy that the DP model also correctly predicts the dipole spiral state to be lower in energy than the single-domain [001] state by 12.1 meV.

\begin{table}[htbp]
\setlength{\belowcaptionskip}{0.05 in}
\caption{DFT absolute energies ($E$ in eV) and relative energies ($\Delta E$ in meV) of four different polar states computed with $1\times1\times15$ supercells. The single-domain [001] state is chosen as the reference for the calculations of $\Delta E$.}
\centering
\begin{tabular}
{c|cccc}
\hline
\hline
  & Dipole Spiral   & [001] & $M_A$ $[uu1]$ & $M_B$ $[11u]$\\
\hline
$E$ (eV) &  $-597.231285$ & $-597.221633$ & $-597.193492$ & $-597.184068$ \\
$\Delta E$ (meV) & $-9.7$ & 0 &28.1 &37.6 \\

\hline
\hline
\end{tabular}
\label{DFTenergy}
\end{table}

The stability of these topological states is fundamentally driven by strain and modulated by periodicity. As shown in Fig.~\ref{FigS0}e, the dipole spiral phases only become energetically favorable over the conventional $T$ phase under sufficiently large tensile strain ($a>3.96~\text{\AA}$). Furthermore, the spiral ground state energy decreases with increasing periodicity $N$. This behavior fits perfectly to the LGD model form of $E=B\cdot\sin^2(\pi/N)+C$ (Fig.~\ref{FigS0}e, inset), providing a robust link between our first-principles results and established LGD theories.

\clearpage
\newpage
\section{Energetics of Competing Spiral Manifolds}

\begin{wrapfigure}[]{r}{0.5\textwidth}
  \vspace{-1.\intextsep}
  \centering
  \includegraphics[width=1\linewidth]{{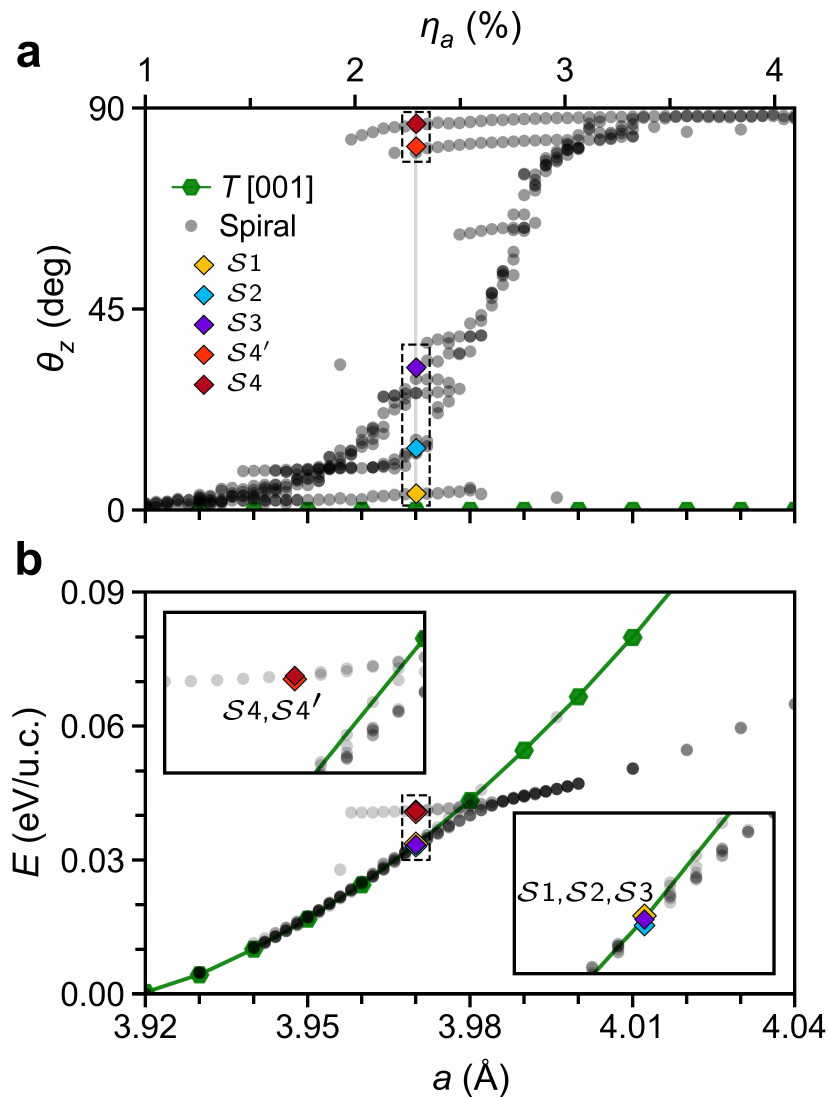}}
  \caption{
  \textbf{(a)} The tilt angle $\theta_z$ as a function of the in-plane strain. The labels $\mathcal{S}$1, $\mathcal{S}$2, $\mathcal{S}$3, $\mathcal{S}$4, and $\mathcal{S}$4$'$ denote distinct final configurations obtained from relaxations with different initial setups.
  \textbf{(b)} Total energy per unit cell ($E$) as a function of the in-plane strain. 
}
  \label{FigS1}
  \vspace{-1\intextsep}
\end{wrapfigure}
Standard structural relaxation algorithms, such as those implemented in the Vienna \textit{ab initio} Simulation Package (\texttt{VASP}), often encounter difficulties in locating complex non-collinear states like dipole spirals. To ensure a comprehensive exploration of the potential energy surface, we therefore performed structural optimizations starting from a wide range of manually constructed initial spiral configurations.

The results of these calculations are summarized in Fig.~\ref{FigS1}, using 1$\times$1$\times$5 pure PbTiO$_3$ supercell. As shown in Fig.~\ref{FigS1}a, different initial setups relax into final states characterized by distinct dipole tilt angles ($\theta_z$) as a function of the applied in-plane biaxial strain. By comparing the total energies of these resultant configurations (Fig.~\ref{FigS1}b), we can distinguish the ground state from metastable states. Our calculations reveal the existence of two separate energy manifolds. The ground state, denoted as the $c$-axis dominated manifold ($\mathcal{M}_1$), comprises three energetically degenerate configurations ($\mathcal{S}$1, $\mathcal{S}$2, and $\mathcal{S}$3). In contrast, the configurations $\mathcal{S}$4 and $\mathcal{S}$4$'$ are also degenerate with each other but possess a higher total energy, thus constituting a metastable, in-plane dominated manifold ($\mathcal{M}_2$). This rigorous approach confirms that a multi-initial-state search is crucial for correctly identifying the complex ground state of the system.

\newpage
\section{Transition Pathways of Helical Dipole Spirals}
We attempted to compute the kinetic energy barriers between these metastable helical configurations using the variable-cell nudged elastic band (VCNEB) method. However, the energy scales involved approached the precision limits of the \texttt{USPEX} algorithm and the helical configuration go out of the spatial search algorithm.

Given the intricate non-collinear nature of the dipole spiral structures, we instead adopted a method based on manual structural interpolation to map the transition pathway. 
This approach is physically justified, as the transformation between helical states ($\mathcal{M}_1  \leftrightarrow \mathcal{M}_2$) naturally proceeds via gradual helical reconfiguration.
Therefore, a path constructed by linearly interpolating the atomic coordinates between the initial and final states provides a physically reasonable approximation of the minimum energy path (MEP), ensuring the reliability of our calculated energy barrier.

\begin{figure}[htbp]
\centering
\includegraphics[width=1\textwidth]{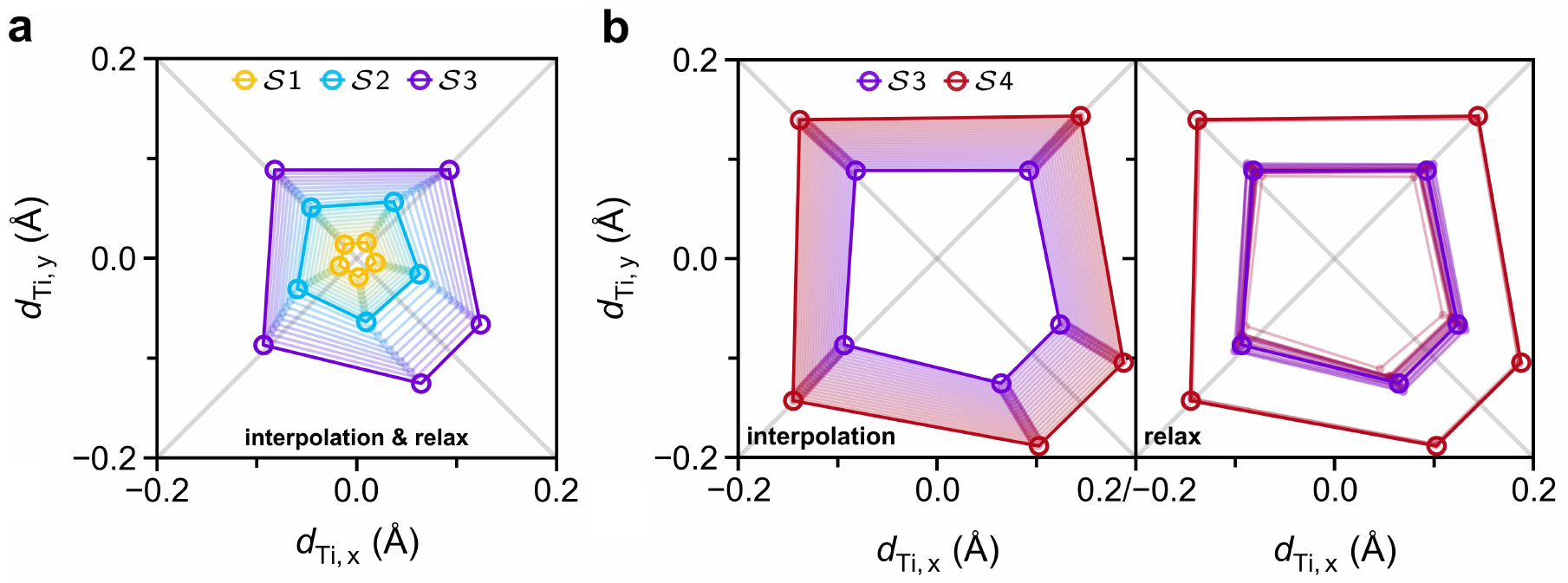}
\caption{
\textbf{(a)} Intermediate states generated by linear interpolation between the three degenerate ground states ($\mathcal{S}$1, $\mathcal{S}$2, $\mathcal{S}$3) of the $\mathcal{M}_1$ manifold. Upon relaxation, these states remain on the continuous energy path, indicating a barrierless transition. 
\textbf{(b)} The left panel displays the unrelaxed transition path constructed by linear interpolation between stable configurations $\mathcal{S}$3 (from manifold $\mathcal{M}_1$) and $\mathcal{S}$4 (from manifold $\mathcal{M}_2$). The right panel shows that after structural optimization, all intermediate states relax back to either the initial or final configuration, confirming the existence of a distinct energy barrier between the two manifolds.
    }
\label{FigS2}
\end{figure}

We investigated the kinetic energy barrier by performing structural interpolations. The results, visualized by tracking the evolution of the in-plane displacement $d_{\text{Ti},xy}$, are presented in Fig.~\ref{FigS2}. This corresponds to \textbf{Fig.~1d} in the main text.
For the ground-state manifold $\mathcal{M}_1$, we generated a series of intermediate structures by linearly interpolating between the three degenerate states ($\mathcal{S}$1, $\mathcal{S}$2, and $\mathcal{S}$3) (Fig.~\ref{FigS2}a). Upon full structural relaxation, these interpolated states showed negligible changes in both their atomic positions and total energies, remaining on the pentagon path connecting the initial states. This outcome provides clear evidence that the $\mathcal{M}_1$ manifold constitutes a continuous, flat energy valley where the system can smoothly transition between degenerate configurations.

We constructed a series of intermediate transition-state configurations (Fig.~\ref{FigS2}b, left panel) by linearly interpolating between the two stable configurations, $\mathcal{S}$3 and $\mathcal{S}$4. These configurations represent a possible geometric pathway for the system's evolution from the $\mathcal{M}_1$ to the $\mathcal{M}_2$. Subsequently, we performed unconstrained structural optimization calculations on these interpolated intermediate configurations. The calculations explicitly demonstrate (Fig.~\ref{FigS2}b, right panel) that these intermediate configurations, upon relaxation, all converge to either the initial $\mathcal{S}$3 or the final $\mathcal{S}$4 configuration, without the emergence of any new stable or metastable intermediate states. This result clearly reveals the existence of a definite energy barrier between the $\mathcal{S}$3 and $\mathcal{S}$4 configurations. The system's evolution must overcome this barrier to complete the transition from the $\mathcal{M}_1$ to the $\mathcal{M}_2$, which strongly confirms that this is a first-order structural phase transition with a real physical energy barrier.

\newpage
\section{Energetic Stability of Coupled Non-parallel Helical Chains}

In principle, any helical polarization chain exhibiting a perfect polygonal distribution within a lattice of $C_4$ symmetry, whether composed of a single or multiple sublattices, is expected to possess a rotational zero-energy mode. However, for the specific case of $\mathcal{M}_1$, our first-principles model calculations reveal that the energetic landscape strongly favors a coupled, non-parallel arrangement over single-sublattice helices. This corresponds to \textbf{Fig.~2f} in the main text.

\begin{figure}[htbp]
\centering
\includegraphics[width=0.6\textwidth]{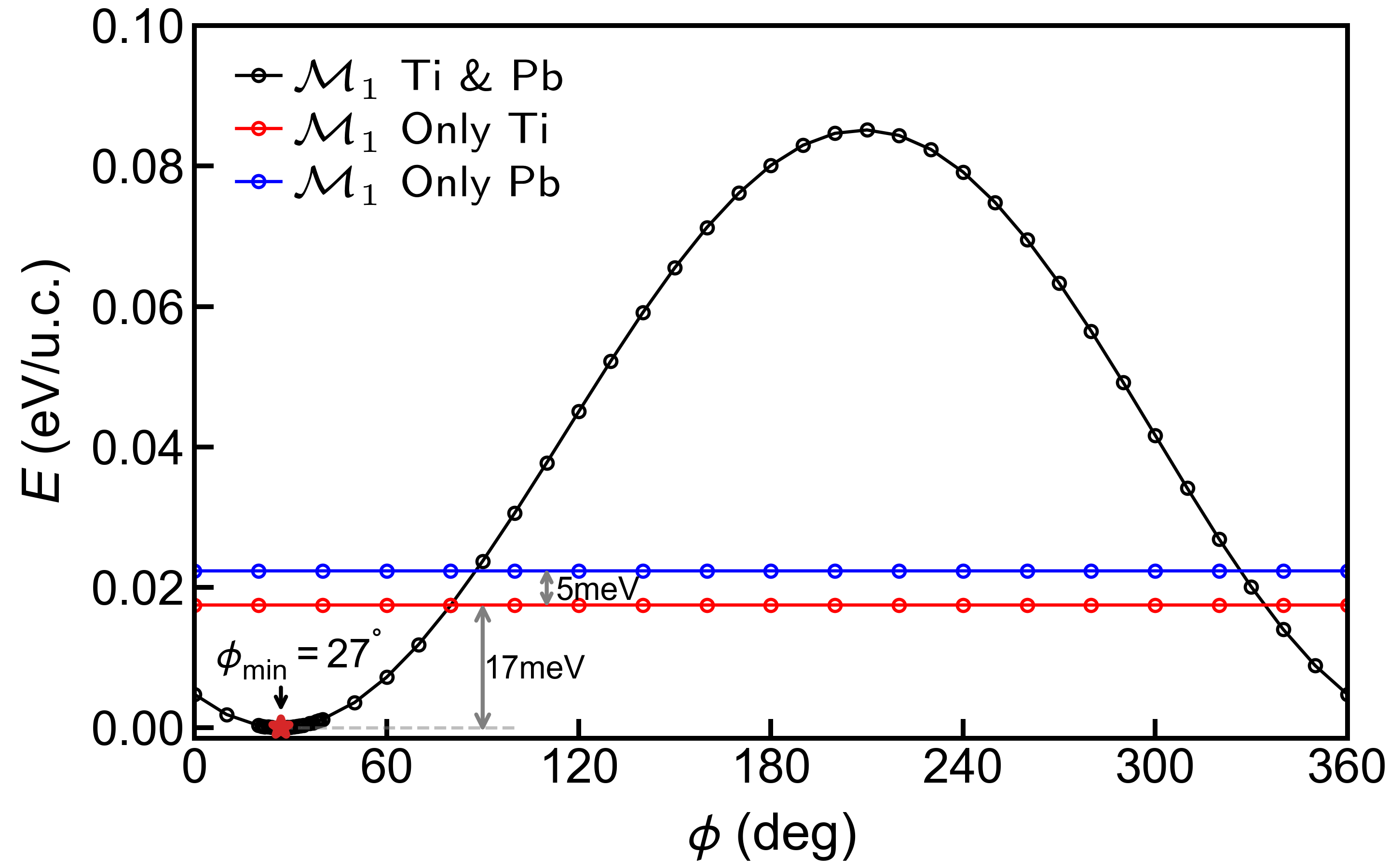}
\caption{
The energy of the coupled Ti and Pb double-chain system (black) is plotted as a function of the relative in-plane angle $\phi$ between the Ti and Pb polarizations. The system has a clear energy minimum at $\phi_{\rm min} = 27^\circ$, establishing this non-parallel arrangement as the ground state (set to $E$=0). For comparison, the energies for hypothetical single-chain helical states, involving only the Ti sublattice (red) or only the Pb sublattice (blue), are shown as horizontal lines. 
}
\label{FigS4}
\end{figure}

As illustrated in Fig.~\ref{FigS4}, we compared the energy of the true ground state (min, red star) with that of hypothetical single-chain helical configurations. A helical spiral formed only by the Ti sublattice is 17 meV/u.c. higher in energy than the ground state. A similar single-chain spiral only on the Pb sublattice is even less stable, with an energy 22 meV/u.c. above the ground state. The true ground state is the coupled, double-chain helical configuration where both Ti and Pb sublattices participate, adopting their preferred non-parallel angle $\phi_{\rm min}$ of $\phi = 27^\circ$. This demonstrates that the inter-sublattice coupling is the dominant factor in stabilizing the helical polar topology, energetically favoring the non-parallel double-chain structure over any single-chain alternative.

\newpage
\section{Evidence for a Soft Rotational Mode}
\begin{wrapfigure}[]{r}{0.54\textwidth}
  \vspace{-1.\intextsep}
  \centering
  \includegraphics[width=1\linewidth]{{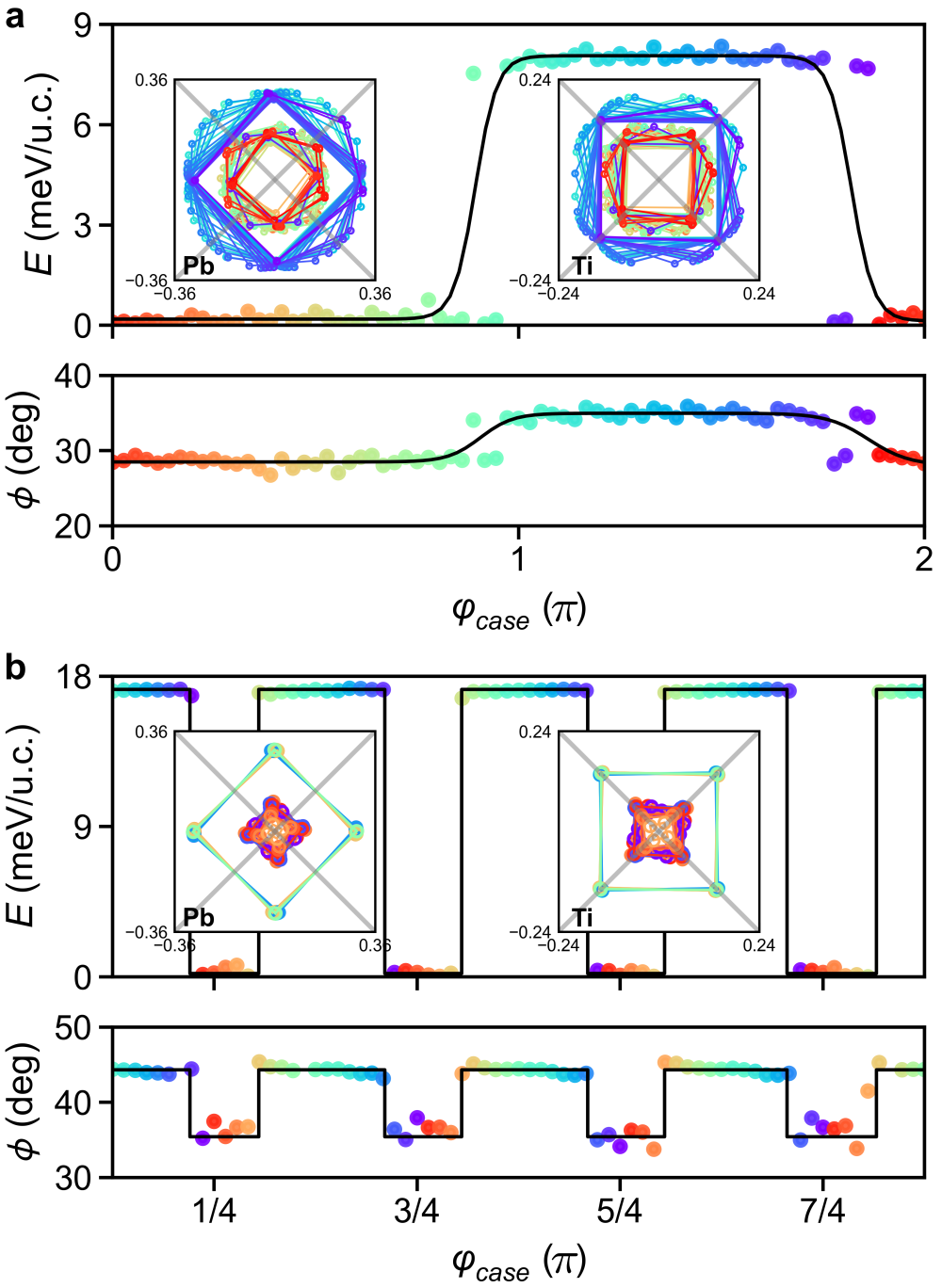}}
  \caption{
      Results are shown for \textbf{(a)} the incommensurate pentagonal model and \textbf{(b)} the commensurate $N$=4 model. Top panels show the energy ($E$) and bottom panels show the relative in-plane angle ($\phi$) between $d_{\rm Ti}$ and $d_{\rm Pb}$, both as a function of the total in-plane polarization phase, $\varphi_{\rm case}$. 
      The insets show the distribution of Pb and Ti atomic displacements in the $xy$ plane. 
}
  \label{FigS3}
  \vspace{-1\intextsep}
\end{wrapfigure}
To investigate the energy landscape of the dipole spiral configurations, we performed structural relaxation calculations starting from idealized, double-helix-parallel states. For our initial calculations on the incommensurate pentagonal model, we constructed configurations where the relative in-plane angle $\phi$ between $d_{\rm Ti}$ and $d_{\rm Pb}$ was initially set to zero. The overall in-plane phase of the total polarization, $\varphi_{\rm case}$, was varied from 0 to $2\pi$. These initial structures were then fully relaxed to obtain the results shown in Fig.~\ref{FigS3}a.

Upon structural optimization, $d_{\rm Ti}$ and $d_{\rm Pb}$ spontaneously relax into a non-parallel state. The calculations reveal two distinct families of helical ground states, corresponding to the $\mathcal{M}_1$ and $\mathcal{M}_2$ manifolds. The $\mathcal{M}_1$ state exhibits a preferred relative angle of $\phi \approx 27^\circ$, while the $\mathcal{M}_2$ state prefers an angle of $\phi \approx 35^\circ$. 
Furthermore, for both manifolds, the in-plane phase of the total polarization $\varphi_{0}$ can be almost continuously varied over the full 0 to $2\pi$ range (Fig.~\ref{FigS3}a, inset). This indicates the presence of a soft rotational mode. 
Due to the structural relaxation, the ideal pentagonal symmetry of the initial model is slightly broken, which destroy the rotational zero-nergy mode. Consequently, rotating the system through the full range of $\varphi_{0}$ incurs a minimal energy cost of approximately 1 meV per unit cell.

To test the generality of this non-parallel coupling, we performed similar calculations for a dipole spiral with a wavelength of $N$=4 unit cells, which is commensurate with the lattice (Fig.~\ref{FigS3}b). Even in this commensurate structure, $d_{\rm Ti}$ and $d_{\rm Pb}$ relax to a non-parallel configuration. We again identify two distinct structural families, one with a preferred angle of $\phi \approx 35^\circ$ and another with $\phi \approx 44^\circ$. 

The persistent emergence of non-parallel polarization in both incommensurate and commensurate models serves as strong evidence for the presence of an electric Dzyaloshinskii-Moriya interaction. This effective interaction drives the A- and B-site polarization chains to adopt a spatially optimized arrangement, resulting in the observed non-parallel helical ground states.

\newpage
\section{Piezoelectric Response of Competing Single-Domain Phases}
\begin{wrapfigure}[]{r}{0.47\textwidth}
  \vspace{-1.\intextsep}
  \centering
  \includegraphics[width=1\linewidth]{{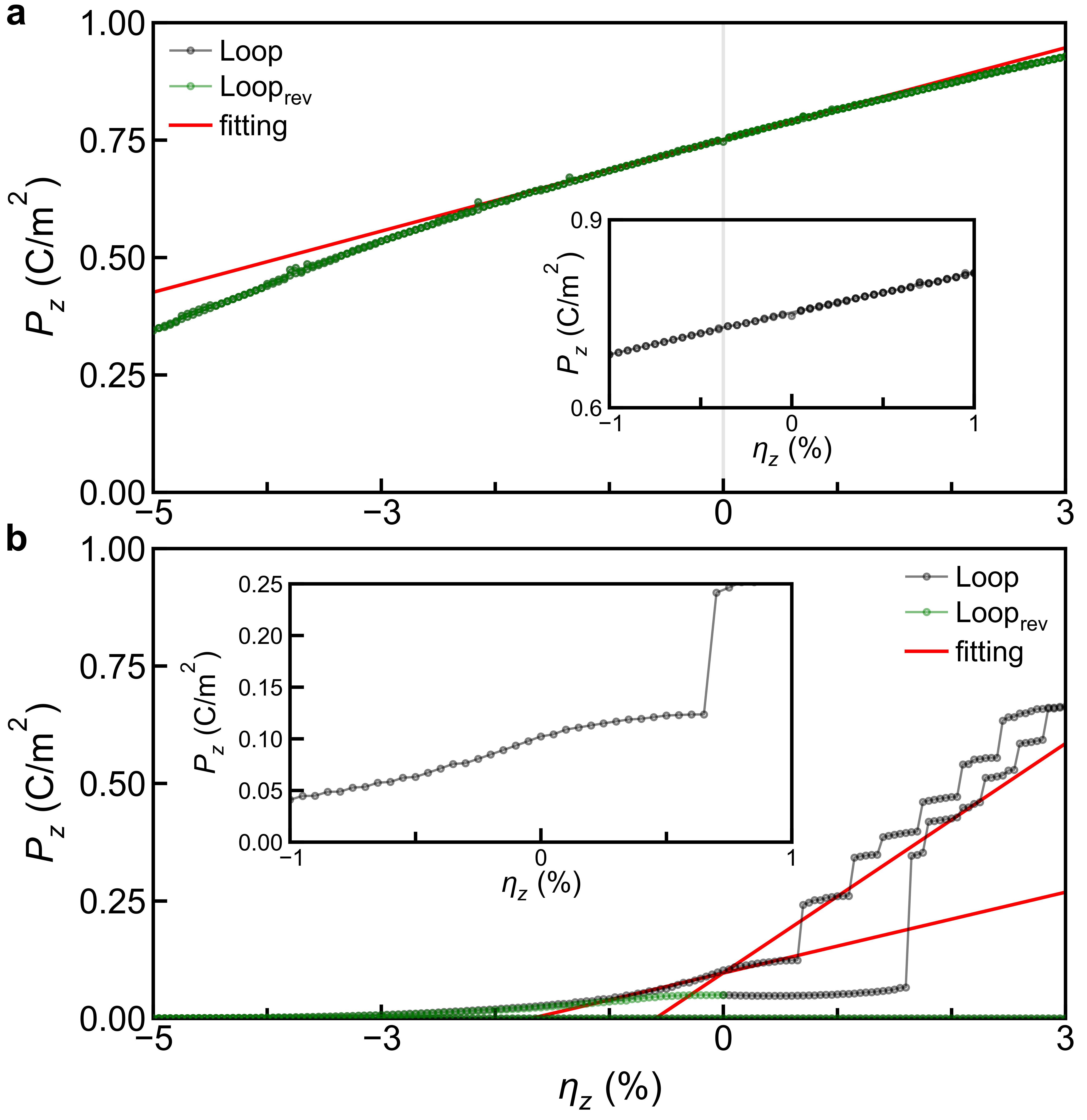}}
  \caption{
    Calculated out-of-plane polarization ($P_z$) as a function of out-of-plane strain ($\eta_z$) for \textbf{(a)} the $T$[001] phase and \textbf{(b)} the $M_B[11u]$ phase. The gray (Loop) and green (Loop$_{\rm rev}$) curves represent the forward and reverse strain cycles, respectively, while the red line is a linear fit.
    The insets show magnified views near zero strain.
}
  \label{FigS5}
  \vspace{-1\intextsep}
\end{wrapfigure}
In the main text \textbf{Fig.~3a}, we presented the piezoelectric response of the dipole spiral structure under a specific in-plane strain ($a$=$b$=3.970 \AA). For a comprehensive comparison, we have also calculated the piezoelectric response of the competing single-domain tetragonal $T$[001] and monoclinic $M_B$[11$u$] (R-like) phases, which can be stabilized under the same in-plane strain condition. The results are detailed below and presented in Fig.~\ref{FigS5}.

The response of the $T$ phase is presented in Fig.~\ref{FigS5}a. The piezoelectric response is almost perfectly linear under weak fields, as highlighted in the inset. The calculated piezoelectric coefficient $e_{33}$, is approximately 5$\sim$6 C/m$^2$. Furthermore, the response remains fully reversible over a wide range of applied strain, from -5\% (compression) to +3\% (tension).

In contrast, the $M_B$ phase exhibits a more complex behavior, as shown in Fig.~\ref{FigS5}b. While the response is approximately linear for very small strains, the overall curve is asymmetric with respect to tensile versus compressive strain, and the effective piezoelectric coefficient is small. Crucially, upon the application of a large out-of-plane compressive strain (-5\%), the $M_B$ phase undergoes an irreversible transformation into a state with predominantly in-plane polarization. This new state remains stable even when the strain is reversed to a large tensile value of +3\%. This irreversible transition indicates that, unlike the dipole spiral, the $M_B$ phase lacks the robust mechanical switchability over a wide strain range. Therefore, it is not a reliable candidate for applications based on mechanical switching.

\newpage
\section{Reproducibility of Mechanical Switching in the Dipole Spiral}

\begin{figure}[htbp]
\centering
\includegraphics[width=0.72\textwidth]{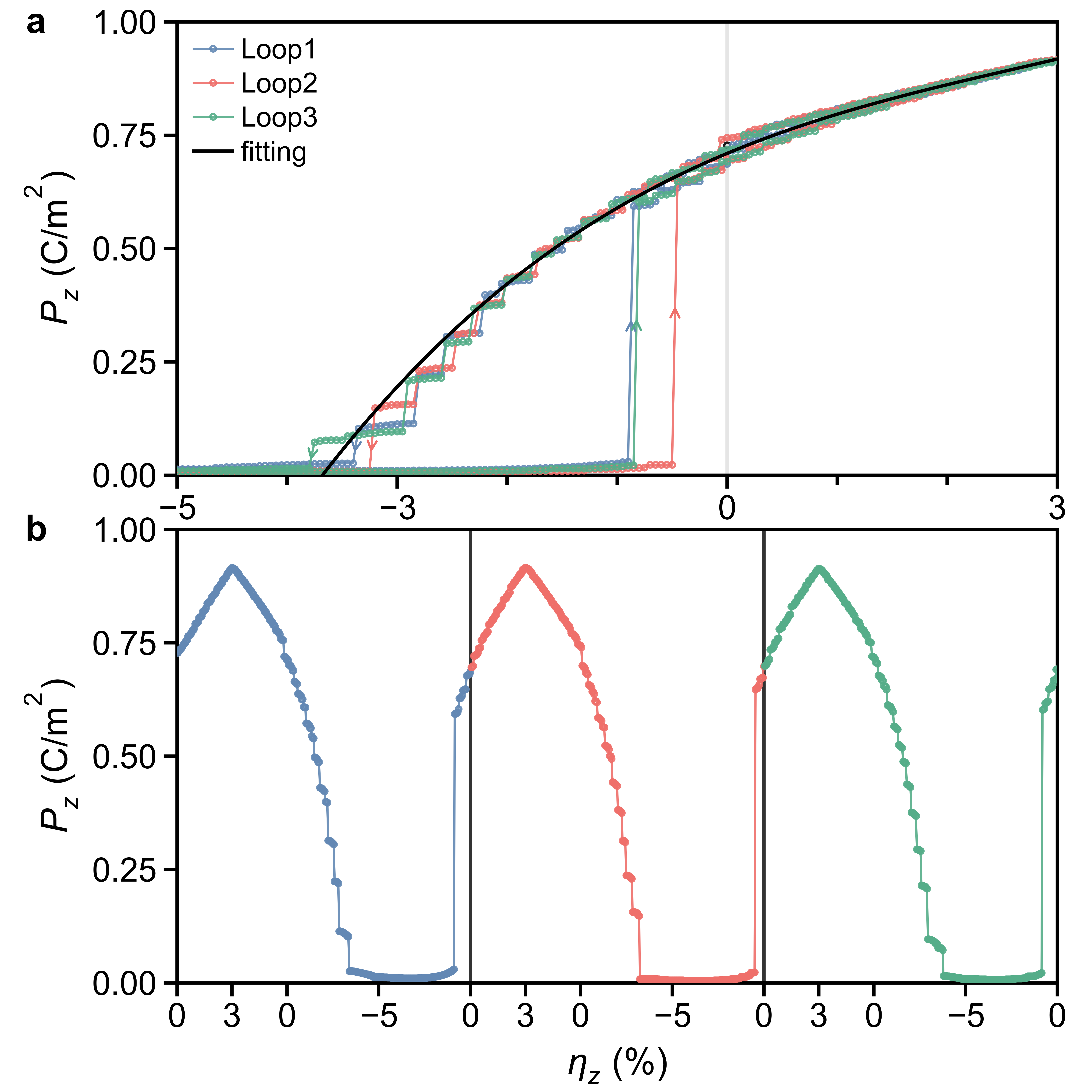}
\caption{
    Calculated out-of-plane polarization ($P_z$) versus out-of-plane uniaxial strain ($\eta_z$) over three consecutive cycles while in-plane lattice is fixed. 
    \textbf{(a)} Superimposed plot of the three loops (Loop1, Loop2, Loop3). Their nearly perfect overlap demonstrates the robust and reproducible nature of the hysteretic, first-order switching. The black solid line corresponds to the anharmonic fit. \textbf{(b)} The same three loops are plotted sequentially to clearly illustrate the cycle-to-cycle consistency.
    }
\label{FigS6}
\end{figure}

\textbf{Fig.~3a} in the main text presents a representative hysteretic loop illustrating the mechanical switching of dipole spiral. To confirm the robustness and reproducibility of this first-order switching, we simulated three consecutive strain cycles. The resulting strain-polarization loops are shown in Fig.~\ref{FigS6}. The nearly identical nature of the three loops underscores the excellent reproducibility of the switching behavior and the structural reliability of the dipole spiral under cyclic mechanical loading.

\newpage
\section{non-volatile hysteretic switching in the Dipole Spiral}

The hysteretic strain-polarization loop presented in the main text \textbf{Fig.~3a} was obtained by incrementally applying an out-of-plane strain, $\eta_z$, while keeping the in-plane lattice parameters fixed, starting from $\mathcal{S}$2. At each strain step, only the atomic positions were allowed to relax. 

\begin{figure}[htbp]
\centering
\includegraphics[width=1\textwidth]{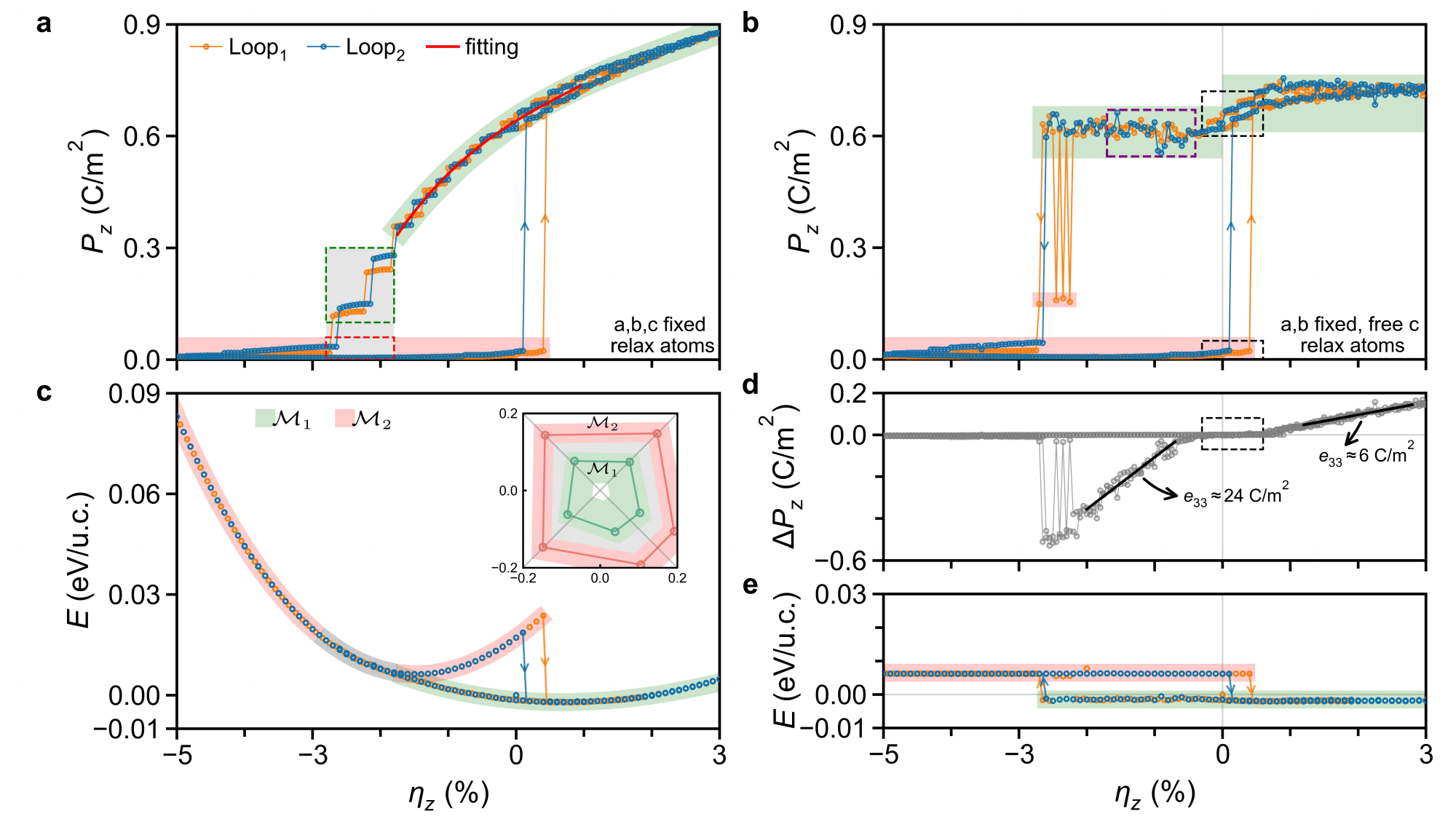}
\caption{
    In the plots, the green, red, and gray shaded regions represent the $\mathcal{M}_1$ manifold, the $\mathcal{M}_2$ manifold, and the intermediate structures stabilized under lattice constraints, respectively.
    \textbf{(a)} The calculated $P_z$-$\eta_z$ loop obtained by fixing the in-plane, out-of-plane lattice and relaxing only atomic positions at each strain step. 
    \textbf{(c)} The corresponding energy landscape for this constrained path. 
    \textbf{(b)} Polarization of the structures after an additional relaxation depend on the configurations in (a) where the out-of-plane lattice parameter is also allowed to optimize. The system collapses onto two distinct manifolds, $\mathcal{M}_1$ and $\mathcal{M}_2$. 
    \textbf{(e)} The energy landscape of the fully relaxed structures, clearly showing the two separate energy branches for the $\mathcal{M}_1$ and $\mathcal{M}_2$. 
    \textbf{(d)} The change in out-of-plane polarization $\Delta P_z$ resulting from the full relaxation process, plotted against the initial out-of-plane strain.
    }
\label{FigS7}
\end{figure}

The $\mathcal{S}$4 spiral is more pronounced, so we initiated the strain cycle from the $\mathcal{S}$4 state to investigate its hysteretic loop.
The resulting polarization and energy profiles for this constrained path starting from $\mathcal{S}$4 are shown in Fig.~\ref{FigS7}a and Fig.~\ref{FigS7}c, respectively.

To investigate the underlying stable energy landscape, we performed an additional structural optimization for each point along this loop. Starting from these constrained structures, we maintained the fixed in-plane lattice parameters but allowed both the out-of-plane lattice constant and all atomic positions to fully relax. As shown in Fig.~\ref{FigS7}b and Fig.~\ref{FigS7}e, the structures converge to two distinct energy manifolds, labeled $\mathcal{M}_1$ and $\mathcal{M}_2$. 
This result provides further verification for the two competing manifolds discussed in \textbf{Fig.~1d} of the main text. 
This result confirms that the $\mathcal{M}_1$ and $\mathcal{M}_2$ states in this system can serve as ``0" and ``1" storage units.
The change in the out-of-plane polarization, $\Delta P_z$, before and after this full relaxation is plotted as a function of the initial out-of-plane strain in Fig.~\ref{FigS7}d.

\clearpage
\newpage
\section{Strain-Tunable Switching in the Hysteretic Cycle of a Dipole Spiral}

\begin{figure}[htbp]
\centering
\includegraphics[width=1\textwidth]{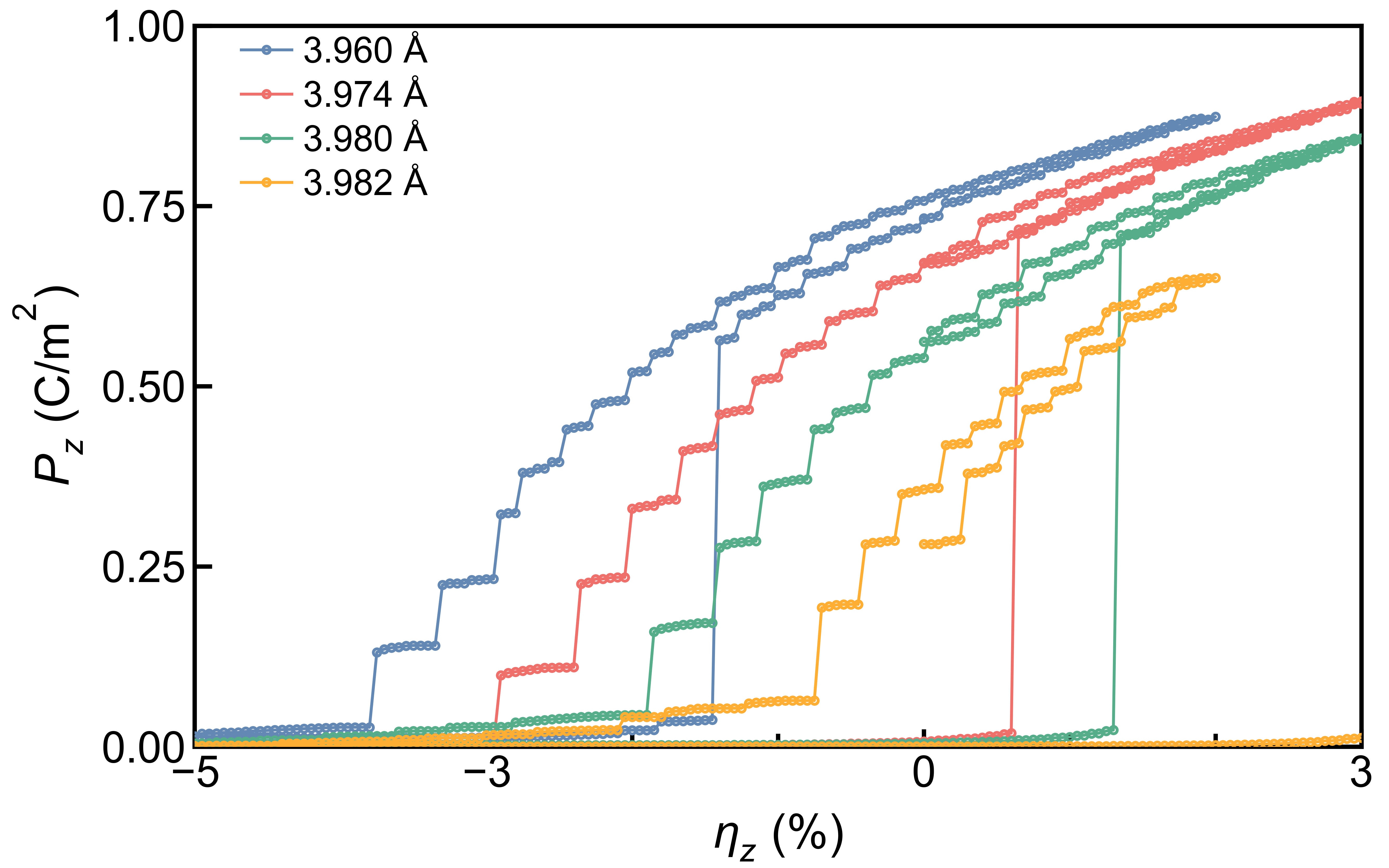}
\caption{Calculated out-of-plane polarization ($P_z$) versus uniaxial strain ($\eta_z$) for the $\mathcal{S}$2 spiral when the in-plane strain is different.
}
\label{FigS7.5}
\end{figure}

The critical tensile strain required for this ``jump back" is not a fixed value but is dependent on the in-plane biaxial strain (Fig.~\ref{FigS7.5}). We can adjust the critical point as needed.
Under large in-plane strain, although the polarization difference between the $\mathcal{M}_1$ and $\mathcal{M}_2$ manifolds is large, the difference in their equilibrium out-of-plane lattice parameters is small. This breaks the reversibility of the transition, causing the hysteretic loop to disappear and trapping the system in a single manifold.  
For example, if the in-plane strain is sufficiently large ($a=b=3.982$ \AA), the energy barrier for the return path becomes insurmountable, and the system remains locked in the low-$P_z$ ($\mathcal{M}_2$) state, preventing the ``jump back" from occurring.

\clearpage
\newpage
\section{Band Structures and Corresponding Projected Density of States}

In the main text \textbf{Fig.~4}, we only presented the band structures and PDOS of dipole spiral for $N$=5 ($\mathcal{S}$4) and $N$=15. Here, we show all the results.
\begin{figure}[htbp]
\centering
\includegraphics[width=1\textwidth]{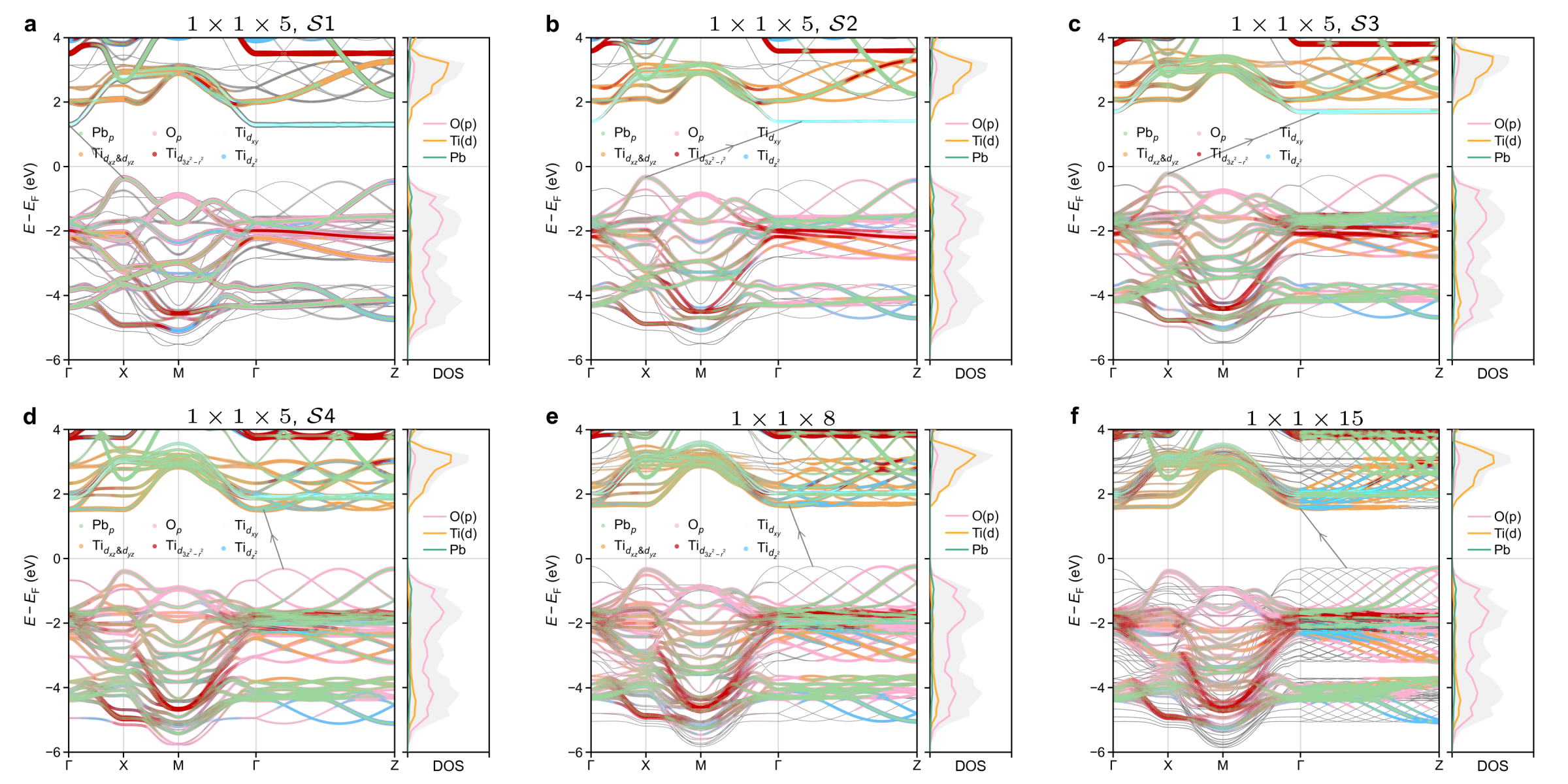}
\caption{
    \textbf{(a-f)} First-principles unfolded electronic band structures and corresponding projected density of states (PDOS) for various dipole spiral configurations. Colors in the bands represent projected contributions from different atomic orbitals. Band structures are shown for spirals with periodicity \textbf{(a-d)} $N=5$ ($\mathcal{S}$1-$\mathcal{S}$4 configurations), \textbf{(e)} $N=8$, and \textbf{(f)} $N=15$.
    }
\label{FigS8}
\end{figure}

\begin{wrapfigure}[]{r}{0.45\textwidth}
  \vspace{-1.\intextsep}
  \centering
  \includegraphics[width=1\linewidth]{{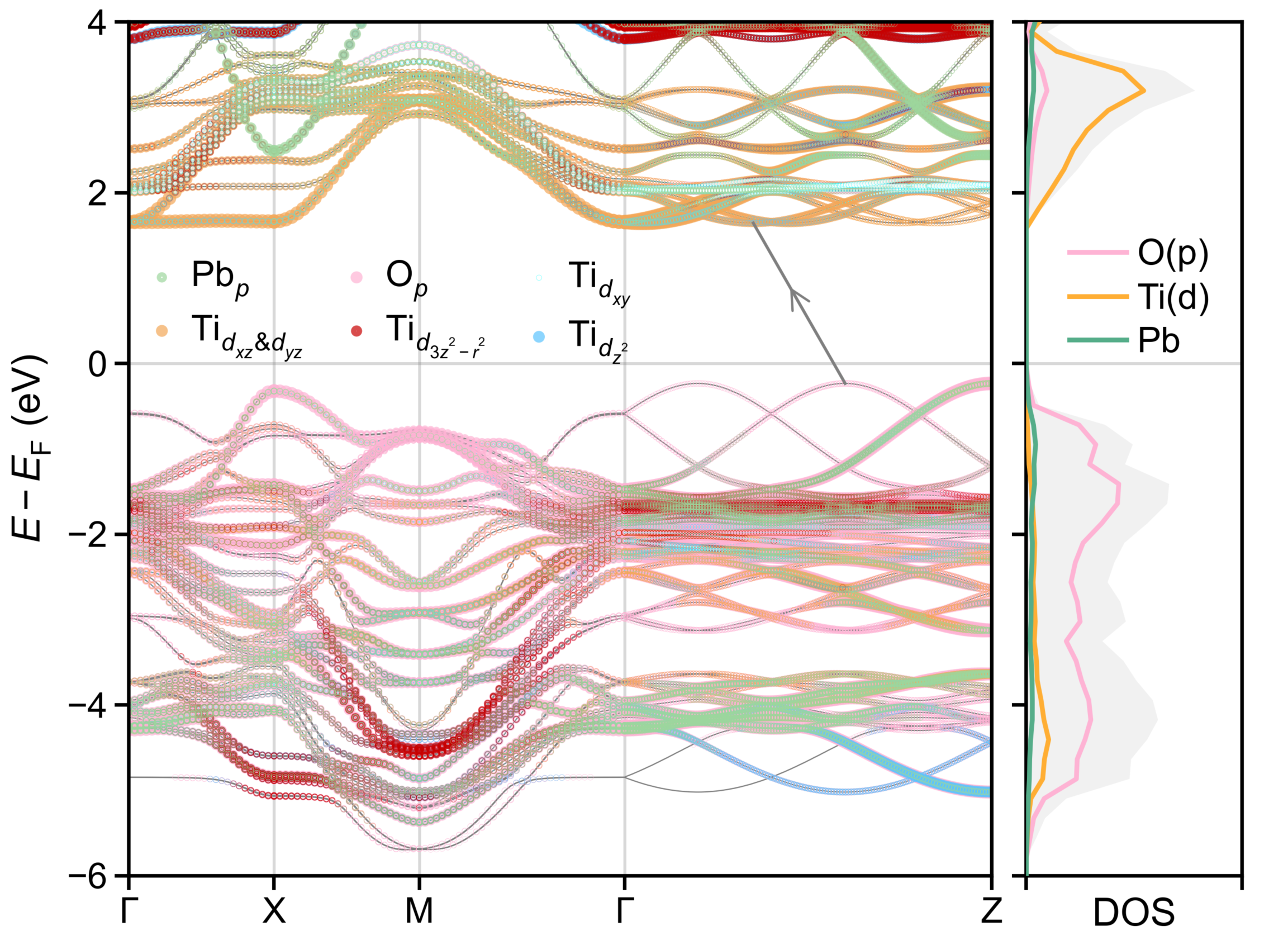}}
  \caption{
    Unfolded electronic band structures and PDOS for the parallel double helix.
}
  \label{FigS9}
  \vspace{-1\intextsep}
\end{wrapfigure}
\vspace{-1\intextsep}
We performed a model calculation using the local Pb and Ti polarizations ($d_{\text{Pb},xy}$, $d_{\text{Pb},z}$, $d_{\text{Ti},xy}$, $d_{\text{Ti},z}$) from the $\mathcal{S}$4 dipole spiral. In this model, we enforced a ``perfect polygon" distribution and set the angle $\phi$ between Pb and Ti helices to zero, representing a parallel double helix. The resulting band structure is similar to that of the fully relaxed $\mathcal{S}$4 state, but more cleaner (Fig.~\ref{FigS9}). This confirms that the woven-shape band structure along the $\Gamma$-$Z$ direction is an intrinsic consequence of the helical arrangement itself.


\clearpage
\newpage
\section*{Appendix}\label{Mathematical proof}


\noindent Proof $\sum_{k=1}^{N}\sin\left(\phi+4\pi k/N\right)=0$ $(N>2, N\in \mathbb{Z} ).$\\
For convenience, we change the summation to run from $k=0$ to $k=N-1$:
\[\sum_{k=1}^{N}\sin\left(\phi+4\pi k/N\right)=\sum_{k=0}^{N-1}\sin\left(\phi+4\pi k/N+4\pi/N)\right) = \sum_{k=0}^{N-1}\sin\left(\phi'+4\pi k/N\right) \]
Using the compound angle formula, we obtain:
\begin{equation}
    \sum_{k=0}^{N-1}\sin(\phi'+4\pi k/N)=\sum_{k=0}^{N-1}\left[\sin(4\pi k/N)\cos(\phi')+\cos(4\pi k/N)\sin(\phi')\right]
    \label{cos}
\end{equation}
Note that the roots of $x^N-1=0$ are:
\[
e^{i {\frac{4\pi\cdot0}{N}}}, e^{i {\frac{4\pi\cdot1}{N}}}, e^{i {\frac{4\pi\cdot2}{N}}},..., e^{i {\frac{4\pi\cdot(N-1)}{N}}}
\]
According to Vieta's formulas which relate the polynomial coefficients to signed sums of products of the roots, it follows that:
\begin{equation}
    \sum_{k=0}^{N-1}e^{i {\frac{4\pi k}{N}}}=0
    \label{plus}
\end{equation}
Similarly, it is easy to show:
\begin{equation}
    \sum_{k=0}^{N-1}e^{-i {\frac{4\pi k}{N}}}=0
    \label{minus}
\end{equation}
The sum of equations~(\ref{plus}) and (\ref{minus}) yields:
\begin{equation}
    0=\sum_{k=0}^{N-1}\left(e^{i {\frac{4\pi k}{N}}} + e^{-i {\frac{4\pi k}{N}}} \right)
    =\sum_{k=0}^{N-1}2\cos(4\pi k/N) \Longrightarrow \sum_{k=0}^{N-1}\cos(4\pi k/N) =0
    \label{plus2}
\end{equation}
while the difference between equations~(\ref{plus}) and (\ref{minus}) gives:
\begin{equation}
    0=\sum_{k=0}^{N-1}\left(e^{i {\frac{4\pi k}{N}}} - e^{-i {\frac{4\pi k}{N}}} \right)
    =\sum_{k=0}^{N-1}2i\sin(4\pi k/N) \Longrightarrow \sum_{k=0}^{N-1}\sin(4\pi k/N) =0
    \label{minus2}
\end{equation}
The substitution of equations~(\ref{plus2}) and (\ref{minus2}) into equation~(\ref{cos}) proves
\begin{equation}
    \sum_{k=1}^{N}\sin\left(\phi+4\pi k/N\right)=\sum_{k=0}^{N-1}\sin(\phi'+4\pi k/N)=\sum_{k=0}^{N-1}\left[0\cdot\cos(\phi')+0\cdot\sin(\phi')\right]=0
\end{equation}

\clearpage

\bibliography{SL.bib}